\documentclass[a4paper,fleqn,usenatbib]{mnras}
\usepackage[T1]{fontenc}
\usepackage{ae,aecompl}

\usepackage{natbib}
\usepackage{epsfig}
\usepackage{graphics}
\usepackage{graphicx}	
\usepackage{amsmath}	
\usepackage{amssymb}	

\def\hbeta{H$\beta$} 
\def\fei{Fe{\small 5015}}
\def\mgb{Mg {\it b}}
\def\feii{Fe{\small 5270}}
\def\feiii{Fe{\small 5335}}

\def\mgfe{[MgFe]'}
\def\fem{$\langle$Fe$\rangle$}
\def\kms{km s$^{-1}$}
\def\mnras{MNRAS}
\def\aap{A\&A}
\def\apj{ApJ}
\def\msun{M$_{\odot}$}
\def\hi{H{\small{I}}}
\def\oiii{[O{\small III}] $\lambda\lambda$ 4959, 5007 \AA}
\def\ni{[N{\small I}] $\lambda\lambda$ 5190, 5200 \AA}

\title[The properties of the bulge and disc of NGC\,3521]{Spectroscopic decomposition of NGC 3521: unveiling the properties of the bulge and disc.}
\author[Coccato et al.]{Lodovico Coccato$^{1}$\thanks{E-mail: lcoccato@eso.org},
Maximilian H. Fabricius$^{2,3}$,
Roberto P. Saglia$^{2,3}$,
Ralf Bender$^{3,2}$, \newauthor
Peter Erwin$^{2}$, 
Niv Drory$^{4}$,
and Lorenzo Morelli$^{5}$
\\
$^{1}$European Southern Observatory, Karl-Schwarzschild-Strasse 2, D-85748 Garching, Germany.\\
$^{2}$Max Planck Institute for Extraterrestrial Physics, Giessenbachstrasse, D-85748 Garching, Germany.\\
$^{3}$University Observatory Munich, Scheinerstrasse 1, D-81679 Munich, Germany.\\
$^{4}$McDonald Observatory, The University of Texas at Austin, 2515 Speedway, Stop C1402, Austin, Texas (USA).\\
$^{5}$Dipartimento di Fisica e Astronomia ``G. Galilei'', Universit\`a di Padova, vicolo dell' Osservatorio 3, 35122 Padova, Italy.\\
}

\date{Accepted XXX. Received YYY; in original form ZZZ}

\pubyear{2017}

\begin{document}
\label{firstpage}
\pagerange{\pageref{firstpage}--\pageref{lastpage}}
\maketitle

\begin{abstract}
  We study the kinematics and the stellar populations of the
    bulge and disc of the spiral galaxy NGC\,3521. At each position in
    the field of view, we separate the contributions of the bulge and
    the disc from the total observed spectrum and study their
    kinematics, age, and metallicities independently. Their properties
    are clearly distinct: the bulge rotates more slowly, has a higher
    velocity dispersion, and is less luminous than the disc.  We
    identify three main populations of stars in NGC\,3521: old
    ($\geq7$ Gyr), intermediate ($\approx$ 3 Gyr), and young ($\leq$1
    Gyr).  The mass and light of NGC\,3521 are dominated by the
    intermediate stellar population. The youngest population
    contributes mostly to the disc component and its contribution
    increases with radius.  We also study the luminosity-weighed
    properties of the stars in NGC\,3521.  Along the photometric major
    axis, we find: i) no age gradient for the stars in the bulge, and
    a negative age gradient for the stars in the disc; ii) negative
    metallicity gradients and sub-solar $\alpha$-enhancement for both
    the bulge and the disc. We propose the following picture for the
    formation of NGC\,3521: initial formation a long time ago ($\geq 7$
    Gyr), followed by a second burst of star formation or a merger
    ($\approx$ 3 Gyrs ago), which contributed predominantly to the
    mass-build up of the bulge.  Recently ($\leq 1$ Gyr), the disc of
    NGC\,3521 experienced an additional episode of star formation
    that started in the innermost regions.
\end{abstract}



\section{Introduction}

Galaxies are generally complex systems that can consist of several structural
components, such as bulges, stellar and gaseous discs, spiral arms, bars, and
shells.  The formation of a galaxy itself and its substructures is a complex
process and can require a number of subsequent events ranging from ``secular
evolution'' (e.g., orbital instability, formation of new stars, quenching) to
``interactions'' (e.g., mergers, accretions, ram pressure stripping); see for
example \citet{Kormendy+04, Naab+17} for reviews.

The detailed study of the various structural components, their
morphology, their kinematics, and their stellar populations can shed
light on the various processes that contributed to their
formation. Because multiple structural components can co-exist in the same
region of a galaxy, such as a small disc engulfed in a large bulge, it is
necessary to remove their mutual contamination from the observations
in order to study them separately. This ``decomposition'' has
been widely applied in the past by fitting analytic functions to the
galaxy light in order to reconstruct the images of the various
components (e.g., \citealt{Kormendy77, Peng+02, Erwin+15}). 
The decomposition can be done also on spectroscopic data, by taking
the different kinematic properties into account. This involves
decomposing the galaxy's line-of-sight velocity distribution (LOSVD)
into multiple kinematic components (e.g. \citealt{Rubin+92,
  Kuijken+93}), and, more recently, decomposing the observed spectrum
into multiple spectral components  (\citealt{Coccato+11},
\citealt{Katkov+11a}, \citealt{Johnston+12}). Key requirements for the
kinematic and spectroscopic decomposition are a relatively good
spectral resolution  and high signal-to-noise,
which are fundamental to properly sample the shape of the LOSVD and
separate the various kinematics and spectral components (e.g.,
\citealt{Fabricius+14, Coccato+14, Coccato+15}).

Recently, we started an observational campaign aimed at studying the detailed
structure of the LOSVDs of the bulges in nearby galaxies, exploiting the
Integral field unit and the superb spectral resolution of the VIRUS-W
spectrograph ($\sigma_{\rm instr} \sim 15$ \kms\ at 5200 \AA) of the McDonald
Observatory. The main goal of the survey is to study the kinematic signatures
of the various components, such as kinematically decoupled cores, and/or a
detailed characterization of the orbital structure that, for example, can be
complex in the case of triaxial potential or bars.

In \citet{Fabricius+14} we studied the spiral galaxy NGC\,7217.
Previous works, including our own, hinted at the existence of two
counter-rotating stellar components \citep{Kuijken+93,
  Fabricius+12}. However, the higher spectral resolution of VIRUS-W
data show that the stars in NGC\,7217 are co-rotating. By studying
the LOSVD, we identified a kinematically hot component (that we
associated with the bulge) and a kinematically cold component (that we
associated with the disc).  A spectral decomposition analysis allowed us
to study the properties of the bulge and the disc in NGC\,7217
independently. Our measurements suggest that NGC\,7217 is in the
process of regrowing a disc inside a more massive and higher
dispersion stellar spheroid.
 
The same analysis was carried on the S0 galaxy NGC\,4191 \citep{Coccato+15}. In
this latter case, we discovered two stellar counter-rotating components, as
suggested by previous studies \citep{Krajnovic+11}, and proposed gas accretion
along two filaments as their formation mechanism.

In this paper, we focus on another galaxy in our sample that has been
claimed to host a stellar counter-rotating component
\citep{Zeilinger+01}: NGC\,3521. NGC\,3521 is a late-type spiral galaxy
with a mixed barred and inner ring morphology, classed as SAB(rs)bc by
the RC3 catalog \citep{rc3}.

One of the first kinematic studies of NGC\,3521 is from \citet{Burbidge+64};
the authors measured the rotation curve of the ionized gas out to $\sim 70$\,kpc
and estimated a total enclosed mass of $8\cdot10^{10}$ \msun\ and an upper
limit of the stellar mass of $2\cdot10^{10}$ \msun. More recent measurements
estimate the stellar and \hi\ masses of NGC\,3521 to be $5\cdot10^{10}$ \msun\
and $8\cdot10^9$ \msun, respectively \citep{Leroy+08, Walter+08}.
\citet{Elson14} studied the kinematics and distribution of the \hi\ halo
surrounding NGC\,3521 out to 25 kpc and found evidence for a secondary \hi\
component. This component, called ``anomalous'' by the author, is distributed
on a thick disc and it rotates more slowly than the main \hi\ component.
Moreover, the spatial distribution of the ``anomalous'' component coincides
with the inner regions of the stellar disc where the star formation rate is
highest; this suggests a link between the stellar feedback and the gas in the
halo, probably driven by galactic fountains. \citet{Elson14} found no evidence
of gas accretion from outside the galaxy system.

\citet{Zeilinger+01} and \citet{Fabricius+12} reported the presence of
a stellar component that is counter-rotating with respect the main
body of the galaxy. The measured velocity separation between the two
stellar components is about 200 \kms. This kinematic decoupling was
interpreted as a ``projection effect induced by the presence of a bar
component seen almost end on'' \citep{Zeilinger+01}. The retrograde
orbits allowed by the bar potential and the viewing angle caused the
bi-modal distribution of line of sight velocities.  This
  ``internal'' origin of the observed kinematic decoupling is
  consistent with the lack of evidence for gas accretion or
  interaction with other galaxies (although it does not prove that
  accretion or interaction did not take place).

The purpose of this paper is to study the kinematics and stellar populations of
the various kinematic and structural components in NGC\,3521, exploiting the
high spectral resolution and sensitivity, and the integral field mode of the VIRUS-W
spectrograph. With the aid of spectral decomposition techniques, we
isolate the contribution of the various structural components from the observed
spectrum and determine their kinematics, age, and metallicity independently in
order to get clues on their formation mechanisms. The advantage of this
approach is that we minimize the mutual contamination of these components in
the regions where their light contribution is comparable.
Moreover, as done for other spiral galaxies with detected stellar
counter-rotation, we test whether or not the presence of a population of
counter-rotating stars is confirmed (e.g. as in NGC\,4191, \citealt{Coccato+15})
or not (e.g., as in NGC\,7217, \citealt{Fabricius+14}) when repeating the
observations with higher spectral resolution. 

The paper is structured as follows. In Sections \ref{sec:fors1} and
\ref{sec:observations} we describe the observations and the data
reduction, in Section \ref{sec:kinematics} we describe the kinematic
measurements, the identification of two kinematically distinct
components, and their association with the galaxy structural
components. In Section \ref{sec:populations} we measure the properties
(age, metallicity, and $\alpha$-enhancement) of the stars of both
components. We describe and discuss the results in Section
\ref{sec:discussion}.  In this paper, we adopt a distance to NGC\,3521
of 8.1 Mpc (as in \citealt{Fabricius+12}), which corresponds to a
linear scale of 40 pc/arcsec.

\section{Photometric observations and data reduction}
\label{sec:fors1}

\begin{figure}
 \psfig{file=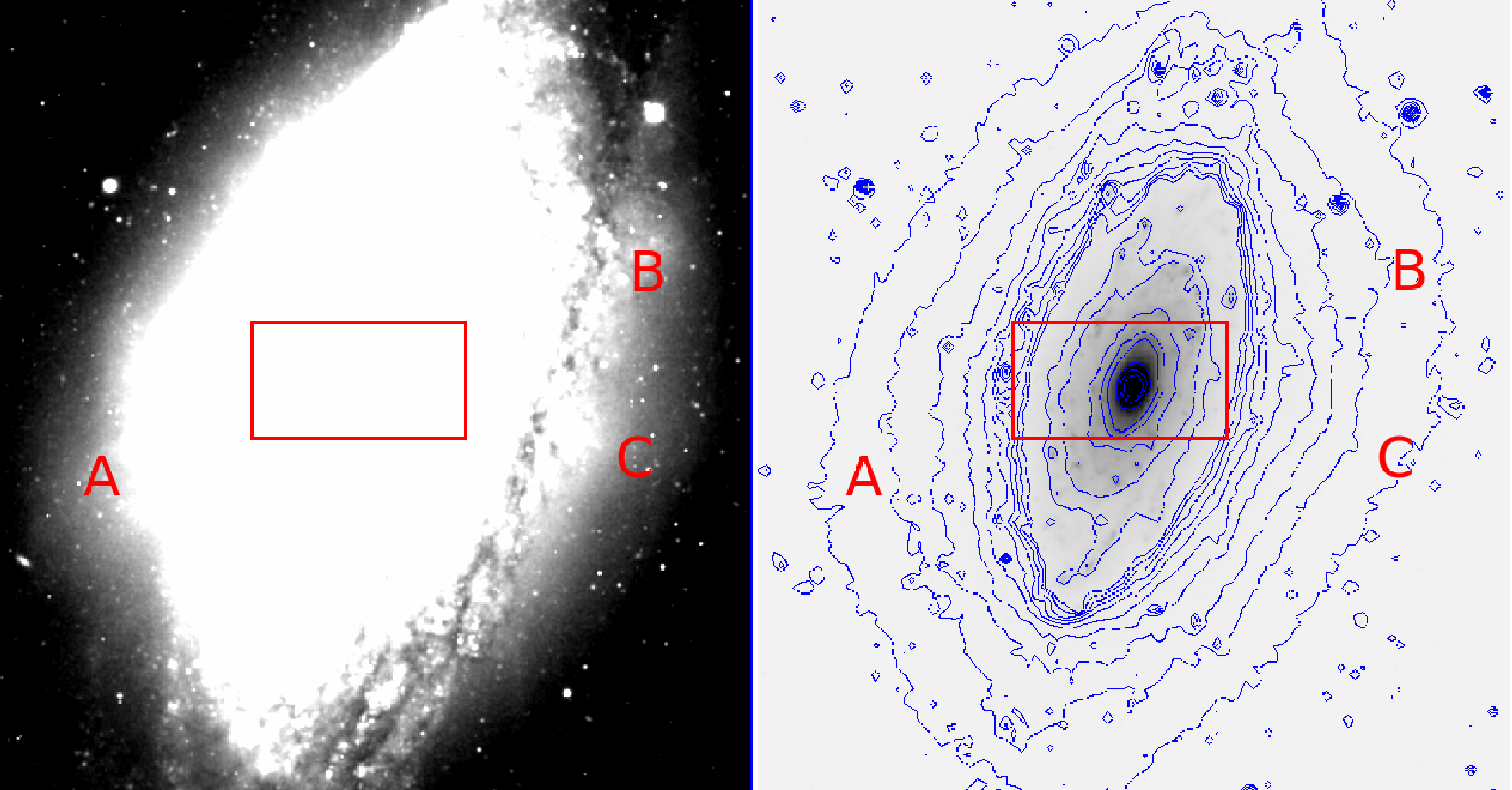,width=8.5cm}
 \psfig{file=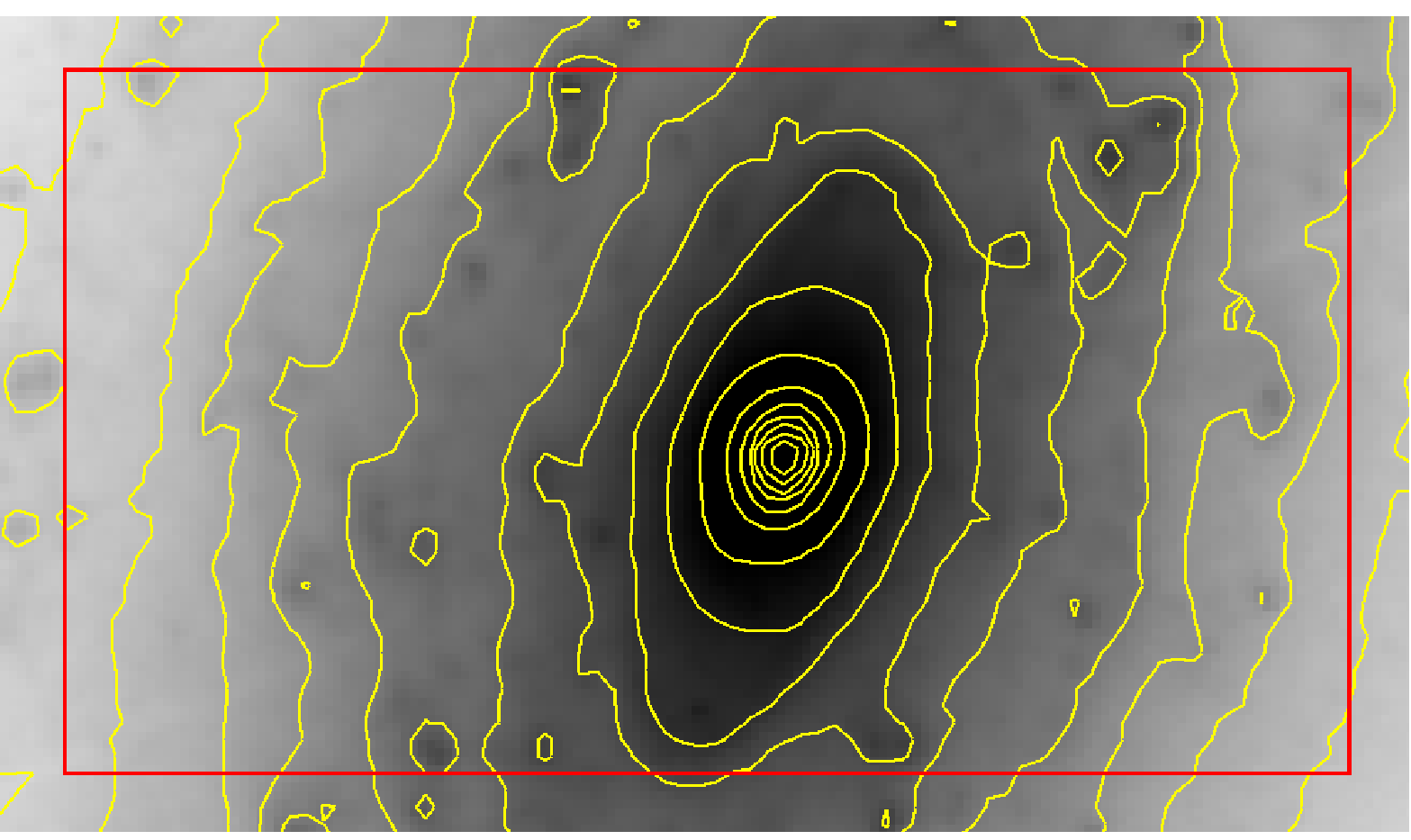,width=8.5cm}
 \caption{VLT-FORS2 B-Band (top left) and Spitzer-IRAC1 3.6
     \micron\ (top right and bottom) images of NGC 3521. The field of
     view in the top panels is $6.\arcmin4 \times 6.\arcmin8$. The
     field of view in the bottom panel is
     $125\arcsec\times 70\arcsec$. Color cuts are selected to
     highlight faint outer regions in the FORS2 image and the inner
     disc and spiral structure in the Spitzer image. Contour plots are
     superimposed to the Spizer image. The three red labels mark the
     position of asymmetric features, that are recognizable as
     ``bumps'' in the light distribution (FORS image) and distortions
     of the isophotes (IRAC image). The red rectangle shows the
     VIRUS-W field of view.}
 \label{fig:fors1}
\end{figure}

We exploit the ESO and {\it Spitzer} archive images of NGC\,3521. 
ESO data were
taken on April $16^{\rm th}$ 1999 with FORS1 at the Very Large
Telescope for the program 63.N-0530 (P.I.  Kudritzki). The galaxy was
observed in the 3 ESO ``Bessel'' filters B, V, and I with a single
exposure of 300 seconds in each filter.  Data reduction (bias and flat
fielding) was performed with the standard ESO FORS pipeline run under
the EsoReflex \citep{Freudling+13} environment and adapting the
available FORS-imaging workflow to handle observations from 1999.
Despite the fact that the exposures were saturated in the centre and the reduced
data are not flux calibrated due to the lack of a standard star
observation, these FORS1 data are deep enough to highlight the
peculiar morphology of the outskirts of NGC\,3521.  The left panel of
  figure \ref{fig:fors1} shows some extended and faint structures
that  depart from an axisymmetric light
distribution (see regions marked with the labels A, B, and C in
  Fig. \ref{fig:fors1}).

 The \textit{Spitzer} IRAC1 (3.6 \micron) image was taken as part as the
  \textit{Spitzer} Infrared Nearby Galaxies Survey
  \citep{kennicutt+03}. The combined mosaic image, with a
  pixel scale of 0.75 arcsec, was produced as part of the SINGS Fifth
  Data
  Delivery\footnote{http://irsa.ipac.caltech.edu/data/SPITZER/SINGS/doc /sings\_fifth\_delivery\_v2.pdf}
  and was retrieved from NED. We performed a final background
  subtraction of the image by measuring median values in approximately
  50 $20 \times 20$-pixel boxes well outside the galaxy and computing
  the mean of those values.

  Because NIR data are less affected by dust absorption, they are
  particularly useful to highlight the inner structure of the
  galaxy. The right panel of Fig. \ref{fig:fors1} shows the inner
  spiral structure of the galaxy and the isophotal contours. Despite
  the RC3 classification as SAB, we found no evidence of a bar. The
  same structures observed in the FORS image are visible in the outer
  isophotes of the Spitzer image, as deviations from an symmetric
  elliptical profile.

The presence of such structure is an indicator that at least one
merging event contributed to the growth of the galaxy in its past.

\section{Spectroscopic observations and data reduction}
\label{sec:observations}
The observations of NGC\,3521 were carried on May 29 and May 30 2011 using the
VIRUS-W Integral Field Unit (IFU) Spectrograph \citep{Fabricius+12} at the 2.7
m Harlan J. Smith Telescope of the McDonald Observatory (Texas, US). The
observed field of view is $105\arcsec \times 55\arcsec$ and it is mapped with
267 fibers of 3\farcs2 diameter on the sky, resulting in a filling factor of 1/3.
Three dithered exposures are sufficient to cover the field of view. We took a
second set of exposures offset by half a fiber diameter with respect to the
first, thereby sub-dithering the observation to increase the spatial resolution.
The individual exposure time was 1800\,s.
Off-set sky exposures of 300\,s each were interleaved with the target
exposures. 
The VIRUS-W instrument was configured in the
high-resolution mode, covering the 4850 -- 5480 \AA\ wavelength range
with a sampling of 0.19 \AA\ pixel$^{-1}$ and a resolving power of
$R \sim 8700$, corresponding to an instrumental dispersion of
$\sigma_{\rm instr} \sim 15$ \kms\ at the centre of the wavelength
range.

Data were reduced using the {\tt FITSTOOLS} package of
\citet{Gossl+02} and the {\tt Cure} pipeline, which was originally
developed for the HETDEX project  \citep{Snigula+14}.  We
followed the prescriptions of \citet{Fabricius+14}. The final reduced
datacube was spatially binned to increase the signal-to-noise ratio
($S/N$) using the implementation for Voronoi Tesselation of
\citet{Cappellari+03} for a total of 83 binned spectra. We
measured\footnote{The computation was done directly on the binned
  spectra by using the {\tt der\_snr} tool available at:
  http://www.stecf.org/software/ASTROsoft/} $40 < S/N < 130$ per pixel
in the spatial bins where the stellar population analysis is performed,
and $20 < S/N < 40$ per pixel elsewhere.

\section{Kinematics and spectral decomposition}
\label{sec:kinematics}

The stellar and ionized-gas kinematics of NGC\,3521 were initially
measured non parametrically in each spatial bin by fitting the
observations with a series of stellar templates. The fitting procedure
recovered the full shape of the line-of-sight velocity distribution
(LOSVD) by exploiting the maximum penalized likelihood method (MPL) of
\citet{Gebhardt2000b} as implemented by \citet{Fabricius+14} to
include \hbeta, \oiii, and \ni\ ionized-gas emission lines in the
fitting process. All the emission lines are set to share the same
kinematics.  MPL determines the LOSVD in a non-parametric way. The
best fitting spectrum is obtained by weighted linear combination of a
number of stellar templates that represent the galaxy stellar
populations; the templates are convolved by the same LOSVD. Therefore,
the algorithm uses the same LOSVD for the different stellar
components.

Then, the recovered LOSVD is modeled as the contribution of two
separate components by fitting two Gaussian functions with independent
velocity, velocity dispersion, and amplitude.  Specifically, we used a
Monte Carlo Markov Chain algorithm based on {\tt pymc} to determine
the maximum likelihood combinations of amplitude, mean velocity, and
velocity dispersion for each of the two Gaussians in each bin.

The knowledge of the kinematics of the two stellar components allowed
us then to disentangle their contribution to the observed spectra in
each spatial bin.  We applied a spectroscopic decomposition technique
\citep{Coccato+11} by using the Python implementation of the Penalized
Pixel Fitting (pPXF) code by \citet{Cappellari+04, Cappellari+17}.  In
contrast with the MPL method, pPXF parametrizes the LOSVD with a
Gaussian-Hermite function. It can handle multiple LOSVDs so that each
stellar component is formed by a linear combination of templates and
convolved with its LOSVD.  In the fitting process, we used the
previously determined kinematics as initial starting parameters in the
fit. This strategy allowed us to obtain the best fitting stellar
spectra for the bulge and the disc for each spatial bin, minimizing
the degeneracy between the fitting parameters and the stellar
components. The kinematics obtained with the pPXF are compatible
within errors to those determined via the LOSVD Gaussian
decomposition.

\begin{figure}
 \psfig{file=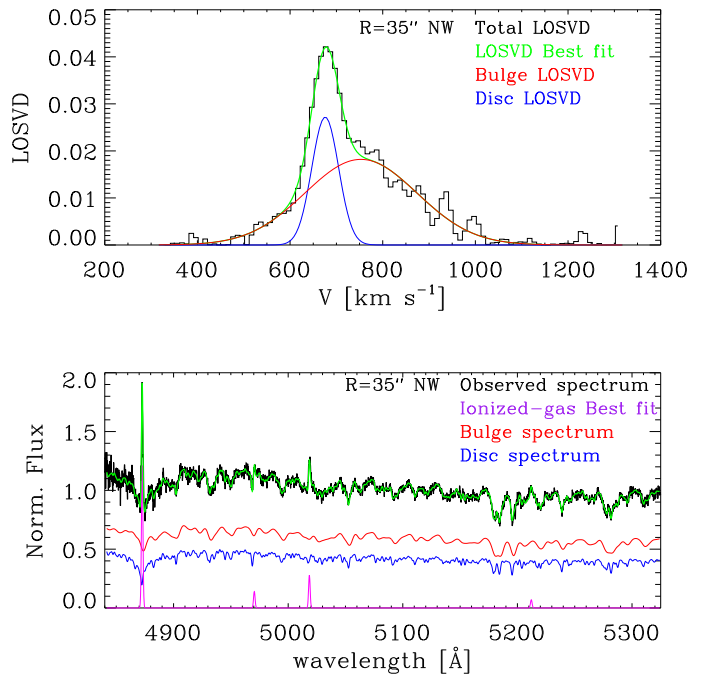,width=9cm,bb=75 360 393 692}
 \caption{Kinematic and spectral decomposition in one spatial bin of
   NGC\,3521. The best fit LOSVD is parameterized by the sum of two
   Gaussian functions (upper panel). The kinematics obtained from the
   LOSVD parametrization is used in the spectral decomposition code to
   extract the spectra of the two components from the observations
   (lower panel). Blue and red represent the kinematically cold and
   kinematically hot components, respectively. Purple shows the
   emission lines and green the stellar best fit models.}
 \label{fig:spectral_decomposition}
\end{figure}

We show in Figure \ref{fig:spectral_decomposition} the
LOSVD and the decomposition for one Voronoi bin along the photometric
major axis.

  The output of the spectroscopic decomposition for each
  Voronoi-binned galaxy spectrum are: i) the kinematics (velocity and
  velocity dispersion of the 2 components); ii) the best-fitting stellar
  templates for each component obtained as linear combination of the
   stars in the input library; iii) the best-fitting stellar models
  of the stellar components, i.e., the best-fitting stellar templates
  convolved for kinematics that account also for the contribution of
  multiplicative polynomials; and iv) the kinematics and best-fit
  model of the ionized gas component.
  The best-fitting stellar templates are used in Section
  \ref{sec:populations} to study the age and metallicity of the two
  stellar components, whereas the best-fitting stellar models are used in
  Section \ref{sec:kinem_results} to study their light distributions.

\begin{figure*}
\vbox{
  \hbox{
    \psfig{file=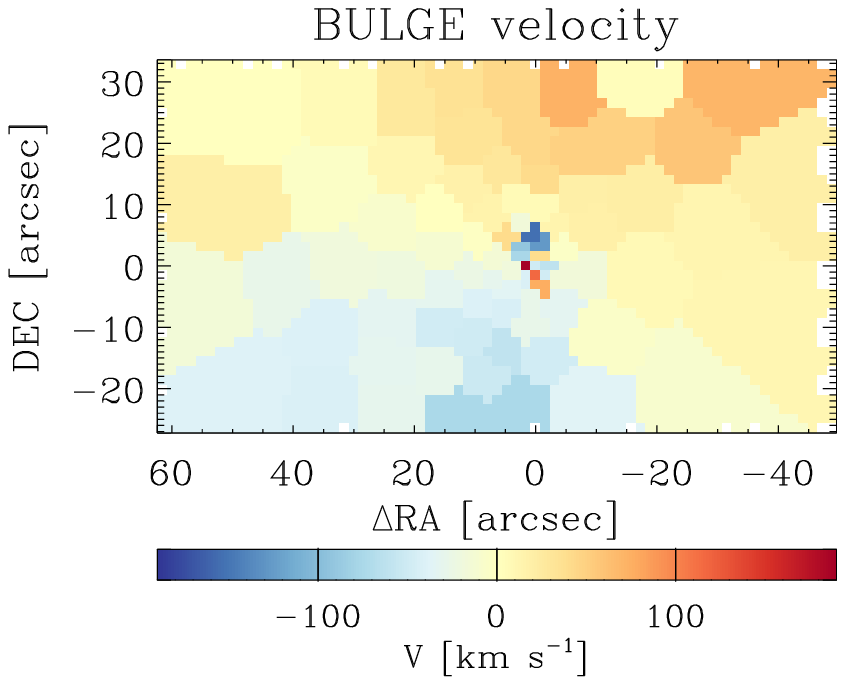,width=6cm,bb=55 340 300 540 }
    \psfig{file=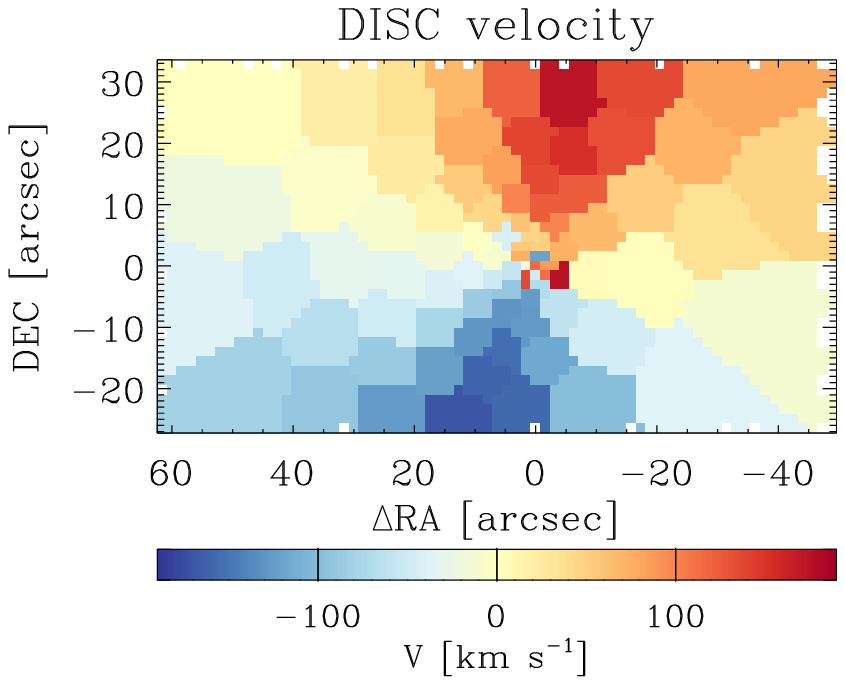,width=6cm,bb=55 340 300 540}
    \psfig{file=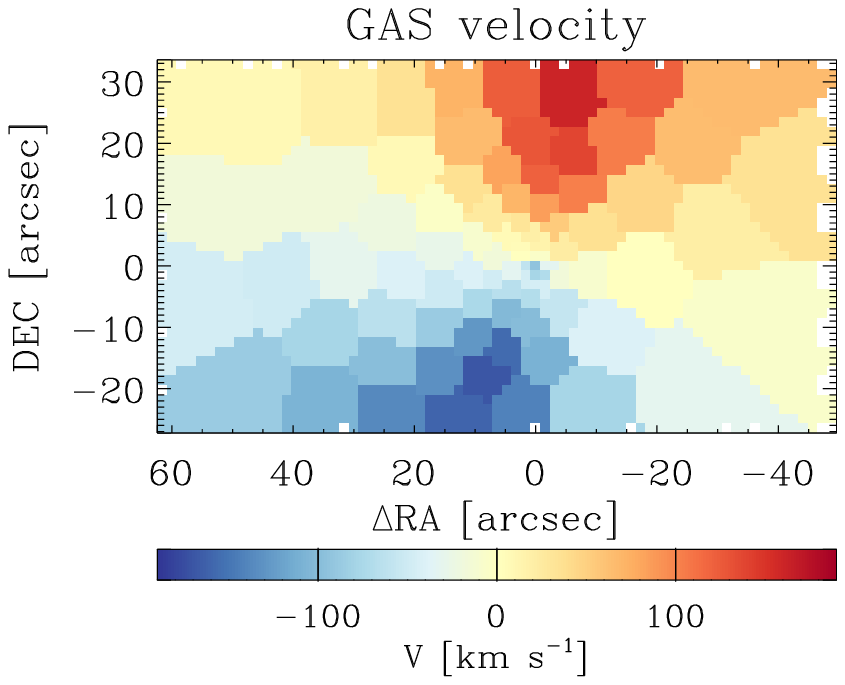,width=6cm,bb=55 340 300 540}}
  \hbox{
    \psfig{file=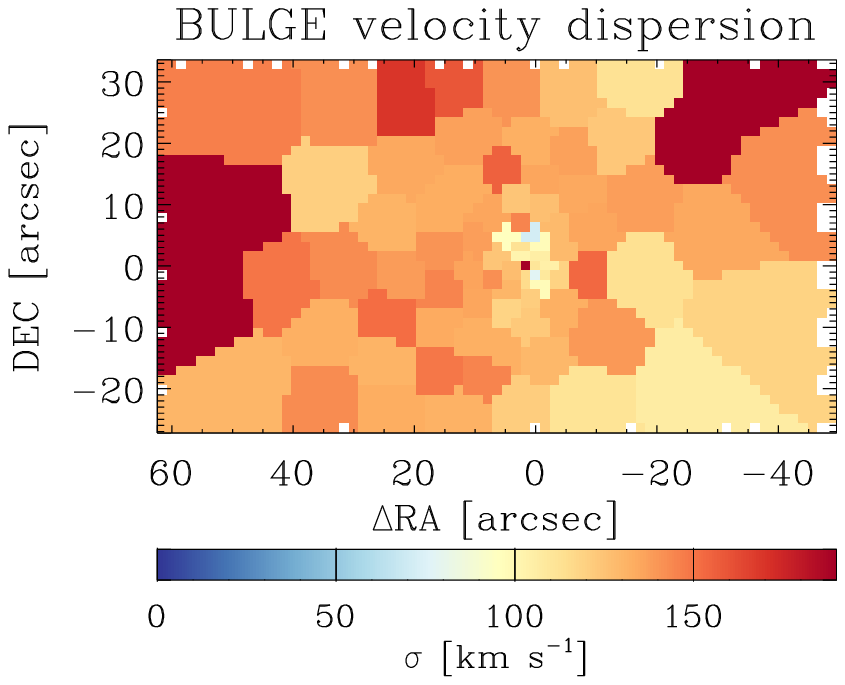,width=6cm,bb=55 340 300 540}
    \psfig{file=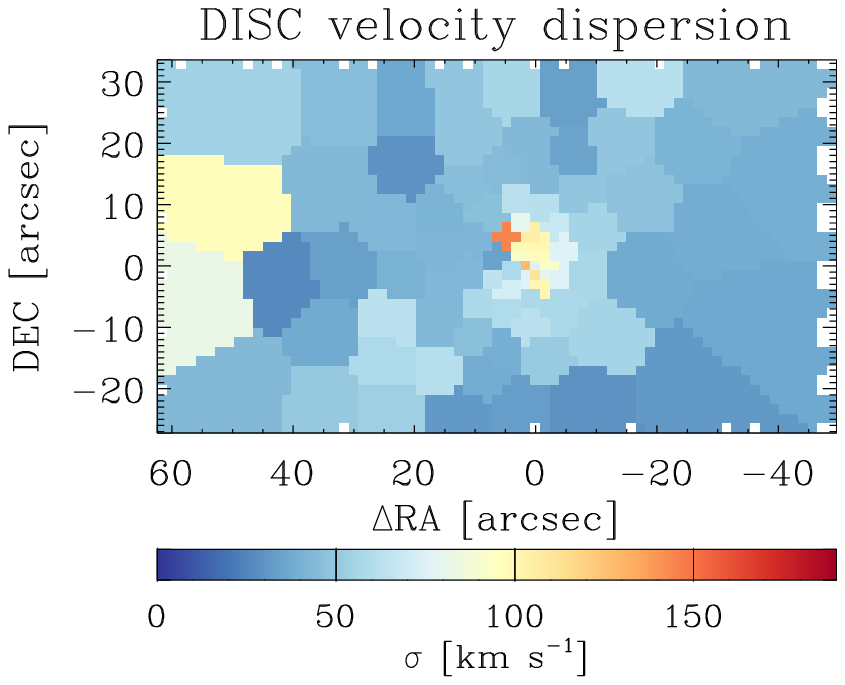,width=6cm,bb=55 340 300 540}
    \psfig{file=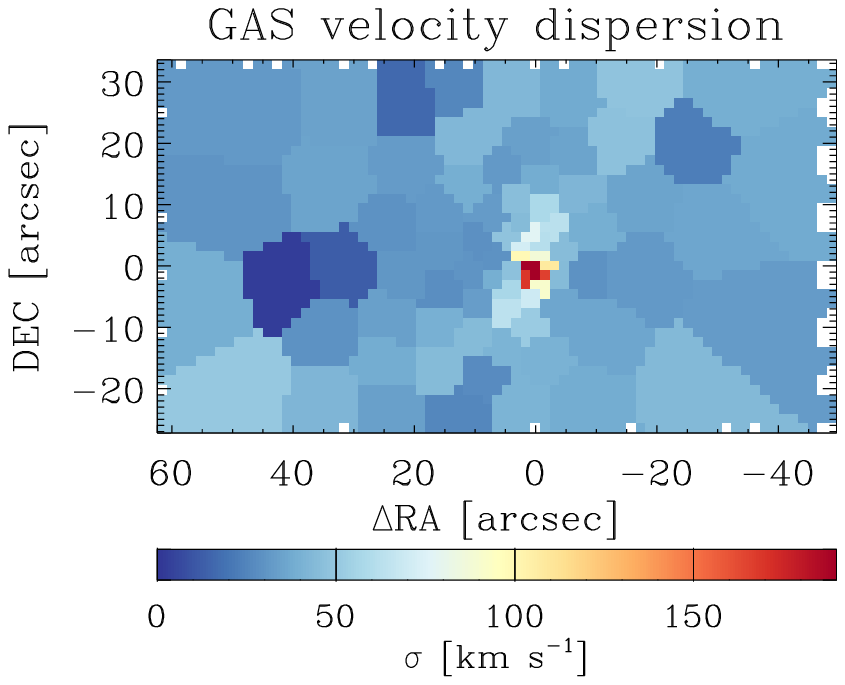,width=6cm,bb=55 340 300 540}}
}
\caption{The two-dimensional velocity (upper panels) and velocity dispersion
(lower panels) fields of the kinematically hot (bulge), the kinematically cold
(disc), and the ionized-gas components in NGC\,3521. North is up, East is left.}
\label{fig:kinematics}
\end{figure*}

\subsection{The extended stellar template library}
\label{sec:library}
To ensure that the kinematic results are not biased by the differences
of the intrinsic properties of the galaxy's stellar population to
those of the template star, it has become common practice to include a
multitude of template stellar spectra in the kinematic extraction
covering a representative region of temperature and metallicity.  In
addition to the line of sight velocities, the spectral decomposition routine 
also determines an optimal set of weights for the linear combination of
templates that ultimately becomes the best fitting model.  Because we
decompose the observed stellar spectra into different components, we
derive stellar population parameters directly from those models rather
than the input data themselves. This further strengthens the need to
include a large set of different stellar types with varying
metallicities and alpha element over-abundances, to prevent a biasing
of the derived stellar population parameters.

From previous work, we already possess a library of spectra for 30
giant stars (A -- M type) observed with VIRUS-W which are also part of
the ELODIE high resolution (R $\simeq$ 42000) library of stellar
spectra \cite{Prugniel+04}. We use these to determine the difference
in spectral resolution between the ELODIE high resolution spectra and
the VIRUS-W spectra, and to then convolve the ELODIE spectra to the
  spectral resolution of VIRUS-W.

\begin{figure}
\psfig{file=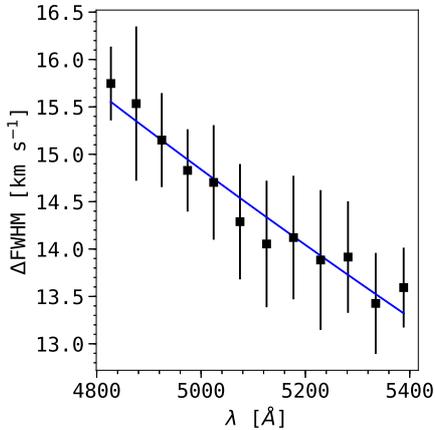,width=6cm}
\caption{
The spectrally dependent broadening function between the 30 VIRUS-W template
spectra and the corresponding ELODIE spectrum.  Each point represents the mean
values of the dispersion that we computed across the 30 spectra and  error bars
represent the RMS in dispersion.  The line corresponds to the linear function
that we use to model the spectrally dependent differential broadening.
}
\label{fig:dsigma_of_lambda}
\end{figure}

The spatial variation of the VIRUS-W's spectral resolution is
negligible ($< 3$\%, \citealt{Fabricius+12}). However, it changes as a
function of wavelength, varying from 17 kms$^{-1}$ around 4800 \AA\ to
14 kms$^{-1}$ at 5400 \AA.  To account for this, we derive a
differential broadening function, which measures as a function of
wavelength the difference in instrumental FWHM between the stars in
the ELODIE library and those observed with VIRUS-W. We initially
divide the spectral range into 12 equally spaced and 50 \AA\ wide
sub-regions and measure the differential broadening separately in
each. For this, we convolve the ELODIE spectrum with a Gaussian and
vary the centroid and the width of the Gaussian until we obtain
minimal residuals between the ELODIE spectrum and the corresponding
VIRUS-W spectrum.  We use the standard \textsc{scipy} least\_squares
routine for the optimization. We repeat this process for each of the
30 spectra that the VIRUS-W library and the ELDOIE library have in
common and then compute the mean and the RMS values in differential
dispersion for each wavelength bin. We find that the change of the
FWHM of the differential broadening with wavelength is well
represented by a first order polynomial (see Figure
\ref{fig:dsigma_of_lambda}); we then use this linear trend as model
for the differential broadening.  We then convolved all of the spectra
of giant stars in ELODIE to the VIRUS-W resolution (as function of
wavelength), by using the linear model of the differential
broadening. The ELODIE high resolution contains spectra of 220 giant
and subgiant stars which is an impractically large number of templates
to use in our kinematic extraction. To fairly sample ages,
metallicities and $\alpha$ over-abundances, we use the published
values for the LICK indices to locate the stars in the four
dimensional space of H$\beta$, Mg\,$b$, $[\mbox{MgFe}]'$, and
$\langle{\rm Fe}\rangle$.  We split this space into one \AA\ wide bins
and select one star per bin, resulting in a total of 72 stars
(broadened to the VIRUS-W resolution) that we use in the further
analysis.

$[\mbox{MgFe}]'$ and $\langle{\rm Fe}\rangle$ are defined by \citet{Thomas+03,
  Gorgas+90}, respectively as:
\begin{eqnarray}
\label{xx}
[\mbox{MgFe}]' & = & \sqrt{\mbox{Mg\,$b$} \cdot \left( 0.82 \cdot {\rm Fe}_{5270} + 0.28 \cdot {\rm Fe}_{5335}\right)}\\
\langle{\rm Fe}\rangle& = & 0.5 \cdot ({\rm Fe}_{5270} + {\rm Fe}_{5335}),
\end{eqnarray}
where ${\rm Fe}_{5270}$ + ${\rm Fe}_{5335}$ are iron Lick indices.

\section{Kinematics and light distribution}
\label{sec:kinem_results}

We found clear evidence of the presence of two distinct stellar
kinematic components in NGC 3521. Our result is supported by two
different methods: the non-parametric LOSVD recovery (MPL) and a
parametric fitting (pPXF). Contrarily to the previous studies
\citep{Zeilinger+01, Fabricius+12}, we found that the two stellar
components in NGC 3521 co-rotate with respect to each other and with
respect to the ionized-gas. The main reason for the disagreement is
that previous data had poorer spectral resolution ($\geq 40$ \kms).
This generated spurious signals in the Fourier space when recovering
the shape of the LOSVD, which was misinterpreted as signal of
counter-rotation.

On the basis of their kinematic properties, we can classify the two
components as kinematically ``cold'' and kinematically ``hot'', as
shown in Figure \ref{fig:kinematics}. As their names suggest, the
``cold'' component has large rotational velocity (up to about 150
\kms\ along the major axis) and low velocity dispersion (about 50
\kms\ over the observed field of view), whereas the ``hot'' component
has low rotational velocity (up to about 50 \kms\ along the major
axis) and large velocity dispersion (about 120 \kms\ over the observed
field of view).  Both the ``hot'' and ``cold'' stellar components show
peculiar kinematic features in the central 3\arcsec, e.g. an apparent
counter-rotating structure in the ``hot'' component velocity field,
irregular velocity field and high values of velocity dispersion in the
``cold'' component. These features are probably artifacts from the
spectral decomposition due to the very small velocity difference
between the stellar components, and not real features.

The ionized gas rotates in the same direction as the ``cold'' stellar
component. Its rotational amplitude is about 150 \kms\ along the major
axis and its velocity dispersion is about 25 \kms\ over the observed
field of view. The ionized-gas velocity dispersion peaks in the
central $\pm 2''$ ($\sigma \sim 100$ \kms); it is not clear if this is
due to unresolved rotation or to the presence of an AGN (NGC 3521 is
classified as LINER, \citealt{Goulding+09}).

It is natural to associate the cold and hot stellar components with
the disc and the bulge of NGC\,3521, despite their identification
being based on kinematics rather than photometry.  In order to further
justify this association, in Section \ref{sec:kinem_phot_results} we
study the light contributions of the two kinematic components, as
returned by the spectral decomposition code, and compare them with those of
the bulge and disc as derived from photometry.

\subsection{Light distribution of bulge and disc}
\label{sec:kinem_phot_results}

In the following sections we study the light distribution of the two
kinematic stellar components identified in Section
\ref{sec:kinem_results}. We follow two complementary approaches: we
study the two-dimensional images (\ref{sec:light_2d}) and the
azimuthally averaged radial profiles (\ref{sec:light_1d}) of NGC 3521
and its structural components.

\subsubsection{Two-dimensional distribution}
\label{sec:light_2d}
\begin{figure}
\psfig{file=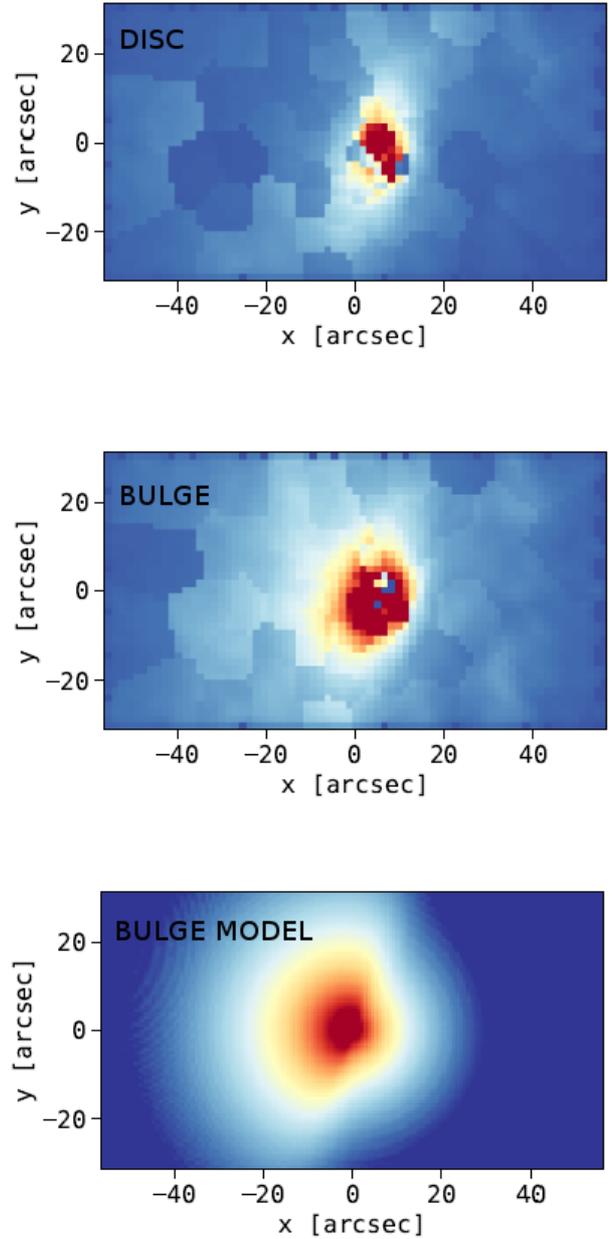,width=8.4cm}
\caption{Reconstructed images of the respective two kinematic
  components of NGC3521 and a model to explain the asymmetry.  The top
  and middle panels are the reconstructed images of the cold and hot
  components, respectively, as they would appear if we could observe
  them separately. The bottom panel shows a projected image of a model
  spheroid that we obtained by a spherical deprojection of the
  photometric bulge by \citet{Fabricius+12}, adding a dust screen
  corresponding in inclination and position angle to the disc of
  NGC3512 and reprojecting this model onto a 2D image.}
\label{fig:decomposition_model}
\end{figure}

Figure \ref{fig:decomposition_model} shows a two-dimensional map of
the light contribution of the two stellar components as derived from
the spectral decomposition.  The map was created in the following
fashion.  First we collapse the VIRUS-W data cube along the spectral
direction to obtain an image on the galaxy.  We then multiply the
integrated flux in each pixel by the relative amplitude that we
obtained from the spectral decomposition code for a given
component. Both the collapsed image and the reconstructed image of the
host stellar component are asymmetric with a higher surface brightness
to the North-Eastern side of the centre as compared to the
South-Western side.  On the other hand the reconstructed image of the
cold component does not show such asymmetry.

This is easily explained by dust obscuration of the light that
originates from the part of the bulge that is located behind the near
side of the disc. The disc itself would -- at least if assumed to be
thin -- appear symmetric in this model because the line of sight
always penetrates about the same amount of disc light before reaching
the dust screen. On the other hand, the bulge is three dimensional:
the side further away will always be more dust obscured than the
closest side, creating an asymmetric image of the bulge.

To better understand this, we executed a simple spherical
decomposition of the bulge light using the decomposition parameters of
\citet{Fabricius+12} with an effective surface brightness of 15.5 mag
arcsec$^{-2}$, an effective radius of $8.5"$ and a S\'{e}rsic index of
3.7.  To model the dust obscuration we then add a infinitely large and
infinitely thin plane through the centre of the spheroid to this model
with an inclination angle of 72.7$^{\circ}$ as derived for NGC3521 by
\citet{Bagetakos+11}. We found that an obscuration to 80\% (in the
sense that 80\% of the light behind the disc is blocked), gives a good
match with the observations. We finally reproject the image onto a two
dimensional plane.

The result is shown in the bottom right of
Fig. \ref{fig:decomposition_model}.  There is reasonable qualitative
agreement between the reconstructed image of the high velocity
dispersion component and the screen spheroid, suggesting that the two
components of the spectral decompositions do indeed correspond to
physical counter parts.

\subsubsection{Radial surface brightness profiles}
\label{sec:light_1d}

\begin{figure}
 \psfig{file=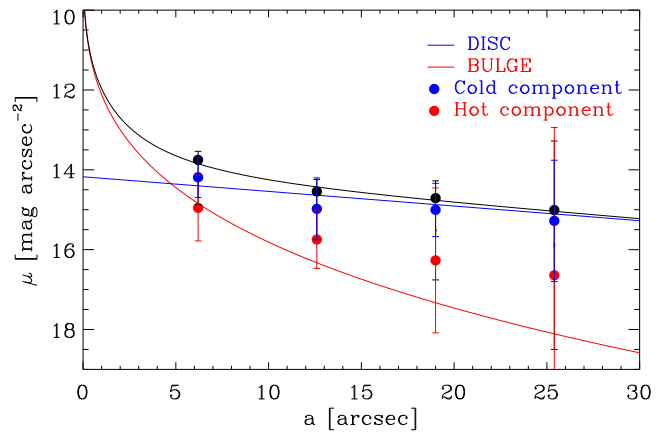,width=8.7cm}
 \caption{Comparison between i) the surface brightness radial
     profiles of the best-fitting photometric model (black line), bulge
     (red line) and disc (blue line) components in NGC\,3521 as
     determined by photometric decomposition (from
     \citealt{Fabricius+12}), and ii) the surface brightness as
     measured on the images of the two kinematic components determined
     by the spectroscopic decomposition (filled circles, this work).
     These images are obtained by first collapsing the VIRUS-W
     datacube along the wavelength direction and by then multiplying
     each of the resulting pixel values by the amplitude of
     the two respective components from the spectral decomposition.
     The black points represent the combined contribution of the two
     kinematic components; their photometric zeropoint is determined
     to minimize the scatter with the best-fitting photometric model
     (black line).}
\label{fig:photometry}
\end{figure}

Figure\ \ref{fig:photometry} shows the bulge/disc surface brightness
profiles as measured by \citet{Fabricius+12} with the -2.5 log of the
median counts measured in elliptical annuli on the hot/cold
components. From Fig. \ref{fig:photometry} we see a qualitative
correspondence, albeit not optimal, between the kinematic and
photometric components. The surface brightness radial profile of the
kinematically cold component well matches that of the disc. The
kinematically hot component has a steeper profile and is less luminous
than the kinematically cold component, in agreement to the bulge
profile. However, the spectral decomposition suggests a somewhat
slower drop-off as function of radius than the photometric
decomposition.  Part of these differences can be attributed to the
difference between the VIRUS-W (optical) spectral range and the
near-IR photometric data.

\subsubsection{Conclusions about the light distribution}

Driven by the similarities described in Sections \ref{sec:light_2d}
and \ref{sec:light_1d}, we refer to the cold and hot components as
``disc'' and ``bulge'' in the rest of the paper. However, the 
differences highlighted in these sections prevented us from
  using the relative light contributions from the photometric
  decomposition as priors for the spectroscopic decomposition.

\begin{figure*}
\hspace{.2cm}
\vbox{
 \hbox{
   \psfig{file=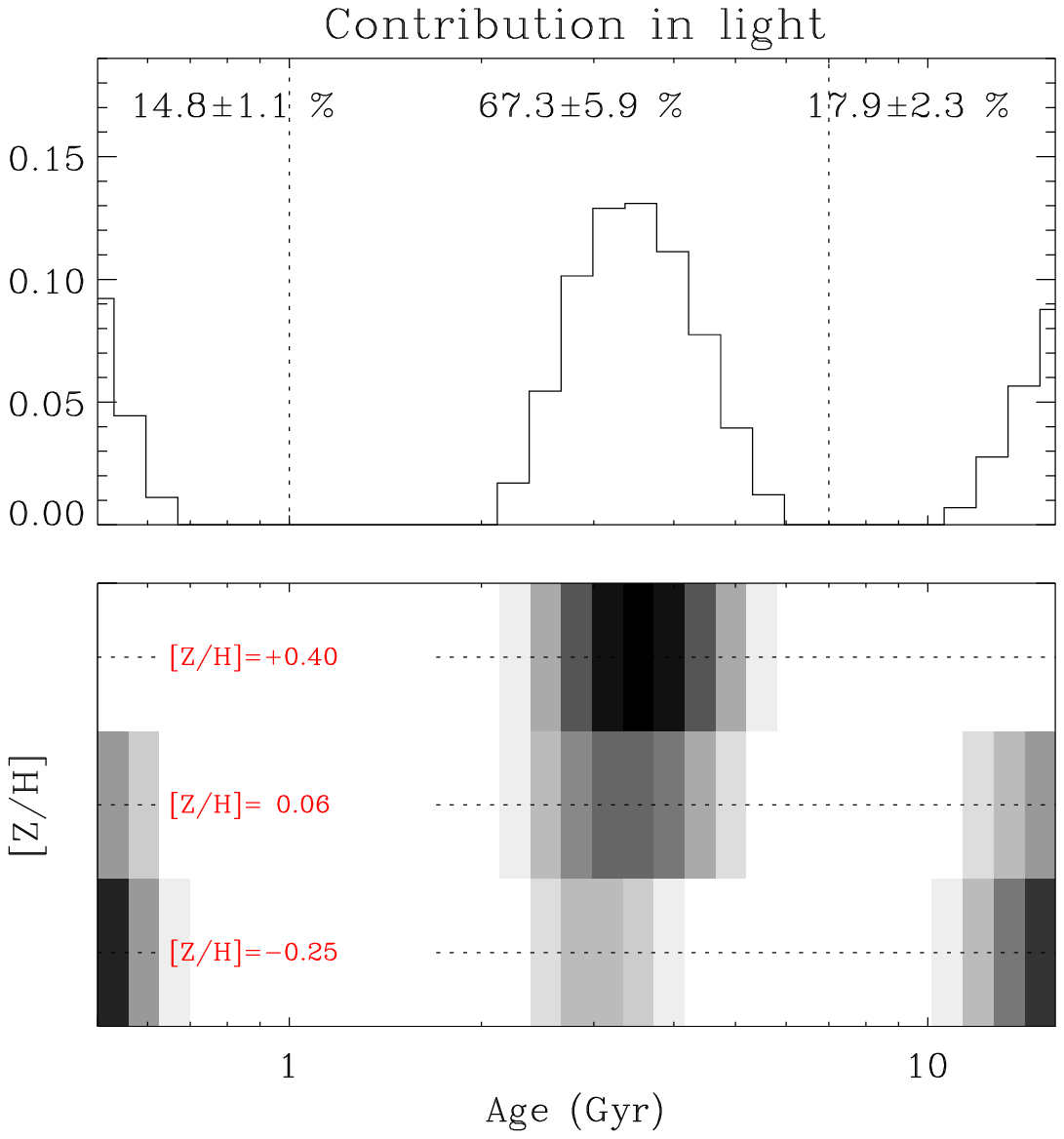,width=7.1cm} 
   \psfig{file=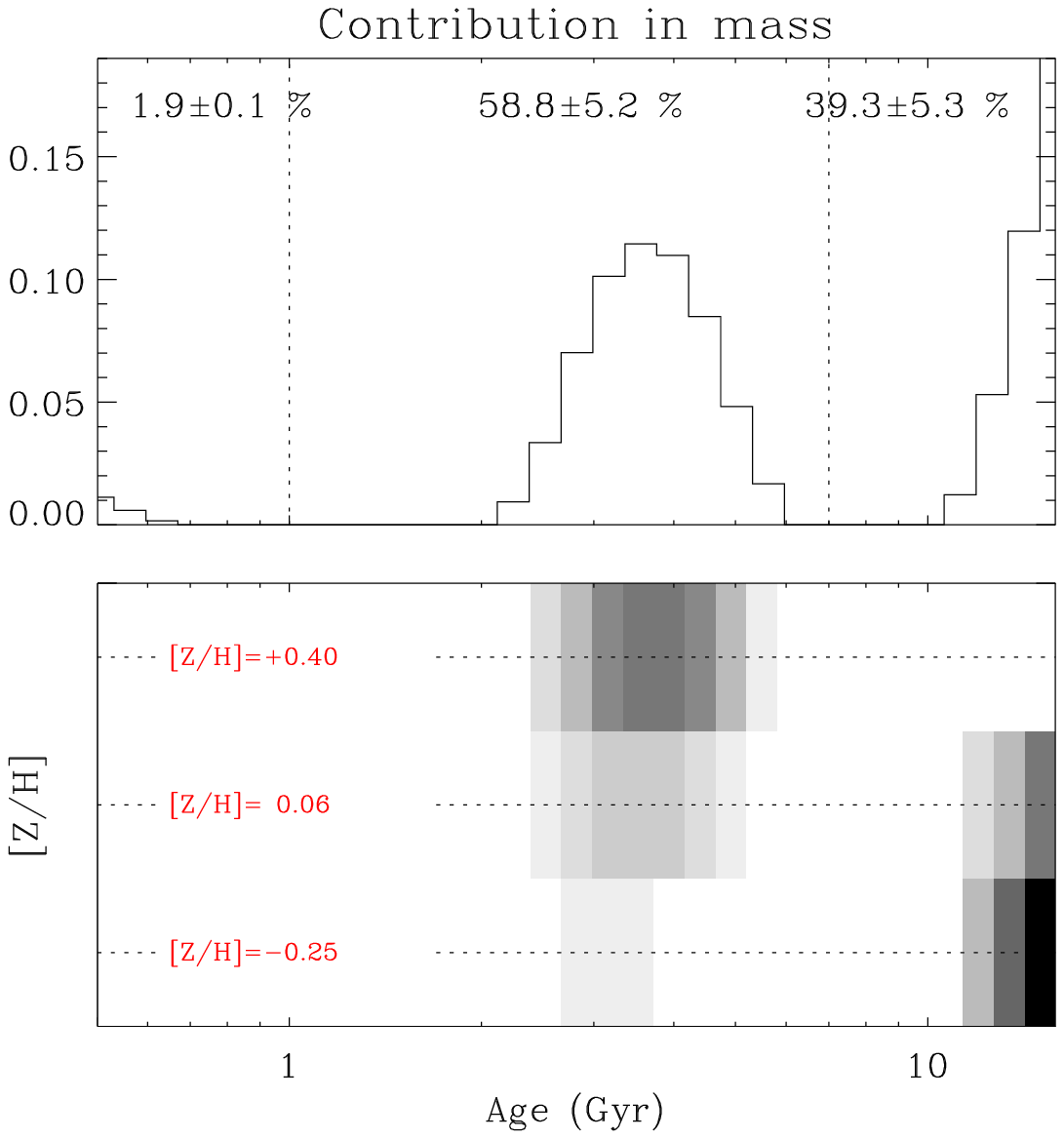,width=7.1cm}} 
 \hbox{
   \psfig{file=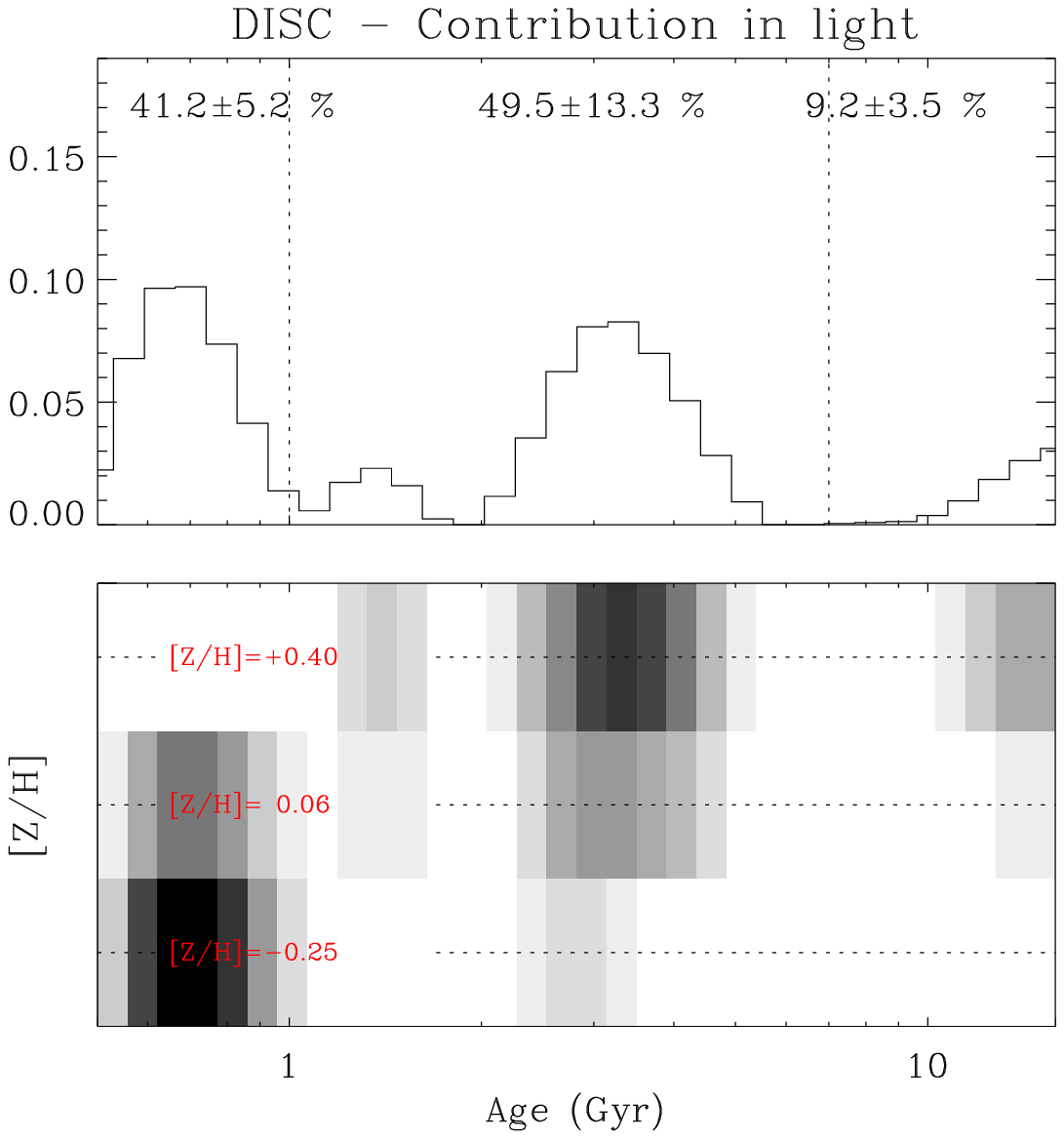,width=7.1cm} 
   \psfig{file=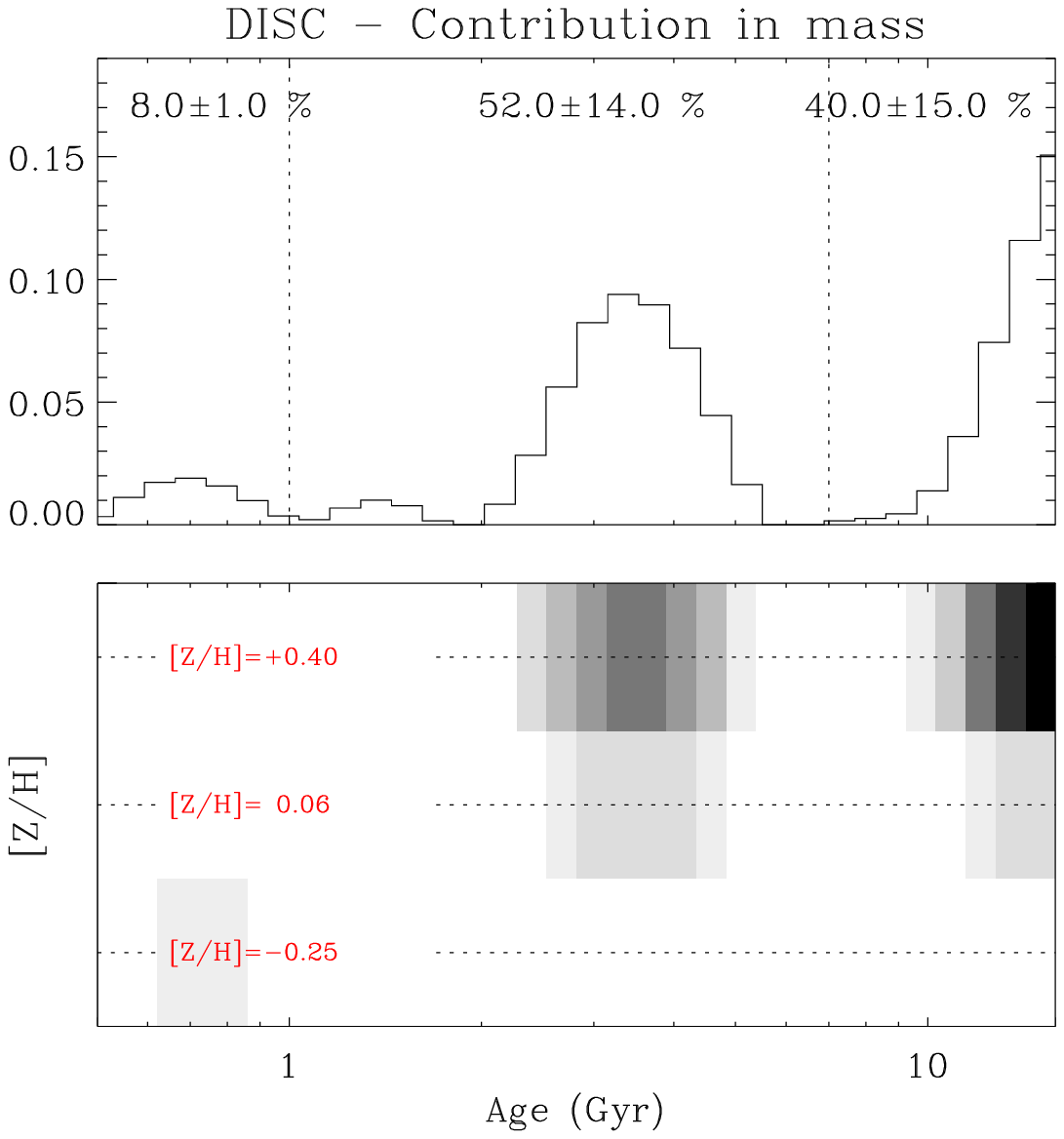,width=7.1cm}}
 \hbox{
   \psfig{file=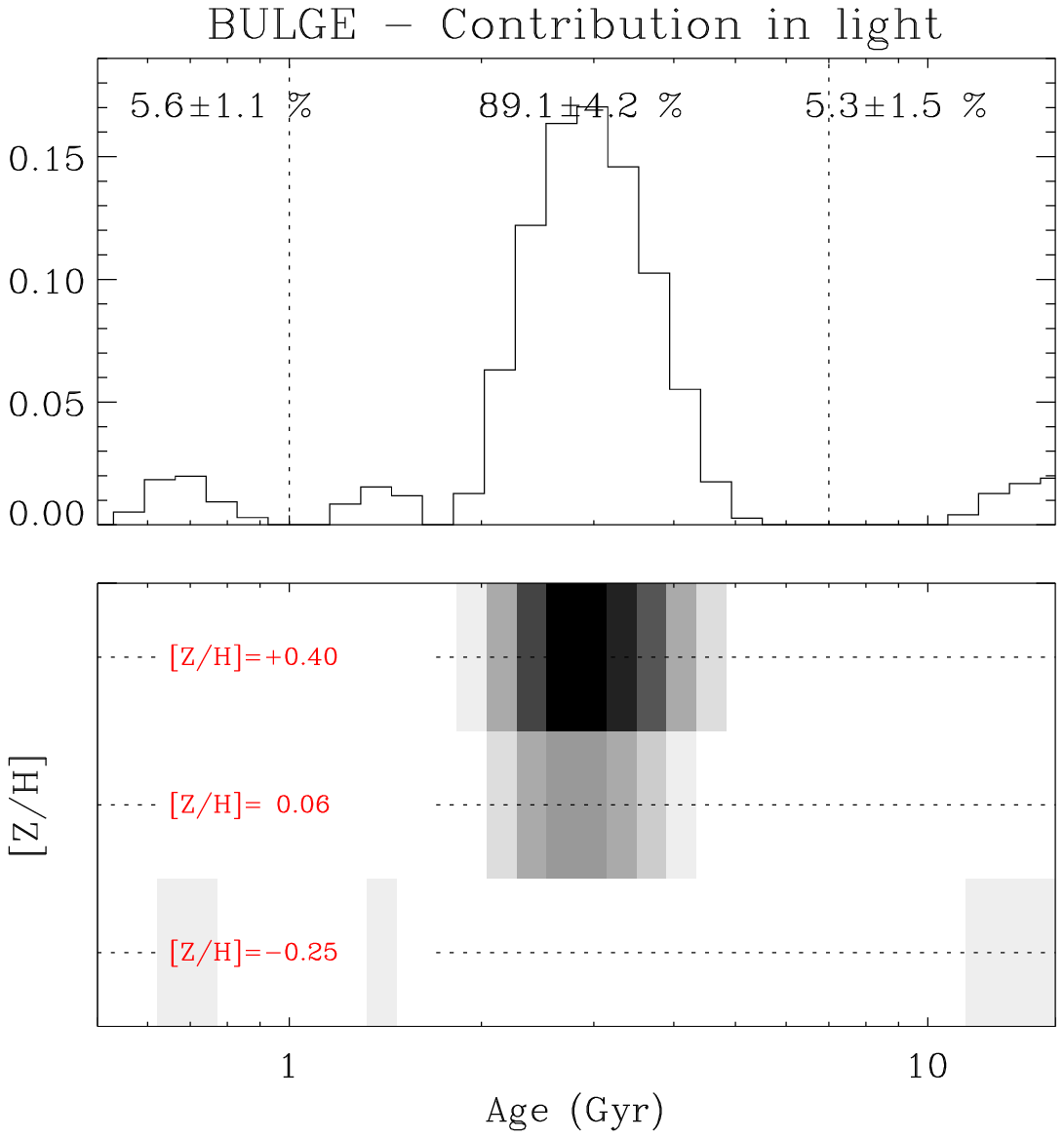,width=7.1cm} 
   \psfig{file=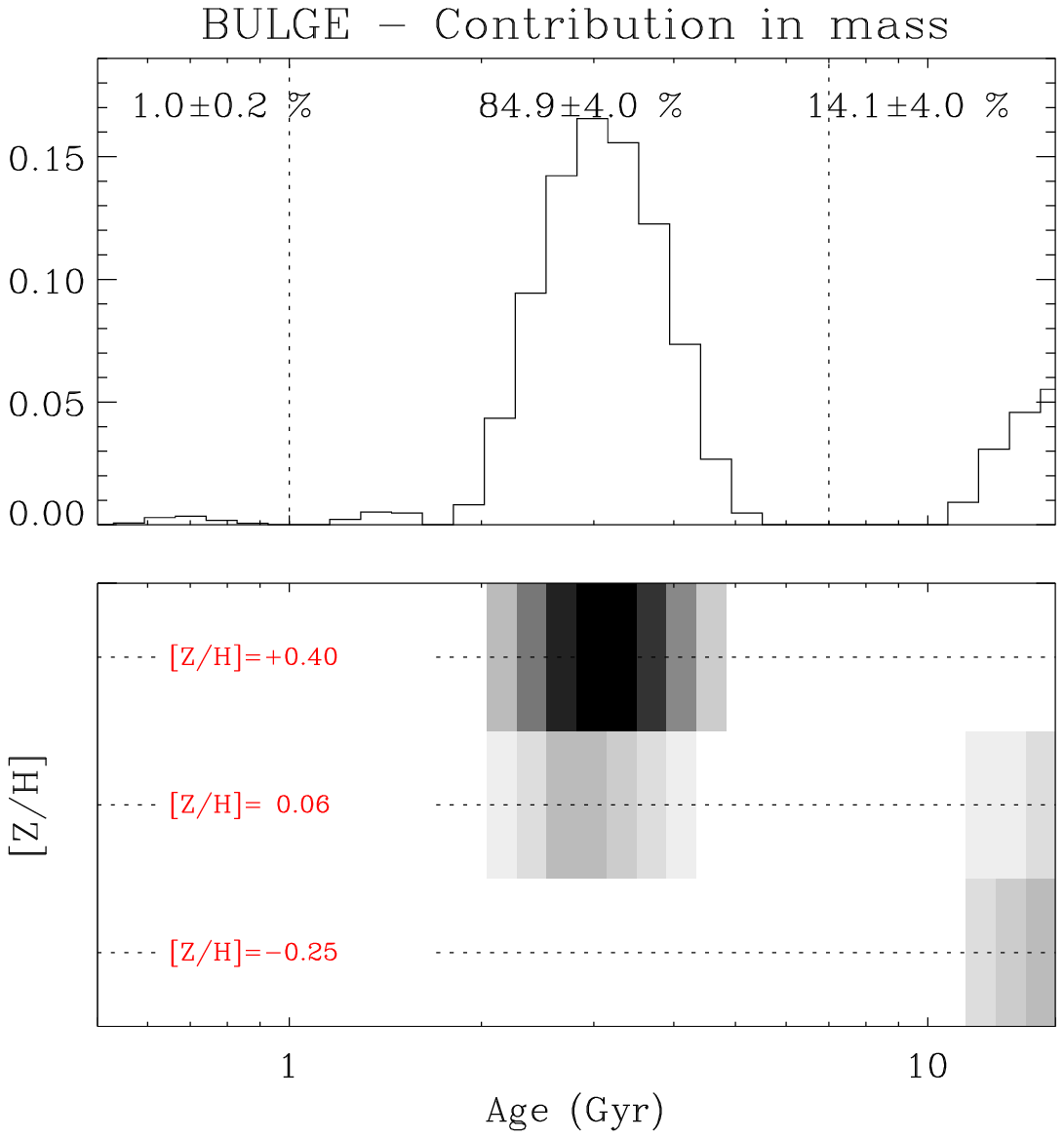,width=7.1cm}}
}
\caption{Contribution in light (left panels) and mass (right panels)
  of individual stellar populations to the total galaxy (top panels),
  disc component (middle panels), and bulge component (bottom panels).
  The histogram values and the gray scales in the upper panels
  correspond directly to the weights that we obtained from pPXF using
  the SSP models.}
\label{fig:sfh}
\end{figure*}

\section{Stellar populations}
\label{sec:populations}

In this section we describe the procedures adopted to study the
properties of the stellar populations of the bulge and disc in NGC
3521 independently. The results will be presented and discussed in
Section \ref{sec:discussion}.  First, we exploit the high
signal-to-noise spectrum obtained by adding all the individual spectra
of each component to derive the luminosity-weighted and the
mass-weighted contributions of the multiple stellar populations that
are present in each component (Section \ref{sec:sfh0}). The analysis
is done by integrating the spectra over the entire radial range
(Section \ref{sec:sfh}) and by considering 3 radial bins to highlight
possible trends with radius (Section \ref{sec:sfh_radial}). Second, we
exploit the information of multiple absorption line indices to derive
the luminosity-weighted simple stellar population (age, metallicity,
and $\alpha$-enhancement) in each spatial bin in order to obtain the
two-dimensional maps of the properties of the bulge and disc (Section
\ref{sec:ssp}).

\subsection{Multiple stellar populations in the disc and the bulge of NGC\,3521}
\label{sec:sfh0}

In this section we identify the presence of multiple stellar
populations in NGC\,3521, by analyzing the observed spectrum and the
spectra of the bulge and disc as determined via the spectroscopic
decomposition. The analysis is done both globally (Section
\ref{sec:sfh}) and in 3 radial bins (Section \ref{sec:sfh_radial}).

In order to minimize the effects of degeneracy from the spectral
decomposition, we considered only the spectra in the bins where the
absolute velocity difference is higher than 50 \kms.  Indeed, we find
that if we decompose the spatial bins where the velocity difference
between the two components is small, the models of the two components
are degenerate: the same total best model is reached when exchanging the
templates from one component to the other. By selecting only bins with
high velocity separation, the degeneracy is minimized by the different
position of the spectral features of the two stellar components.

\subsubsection{Integrated properties}
\label{sec:sfh}

The analysis is divided into 2 parts: first we analyse the observed
spectra of NGC 3521, secondly we analyse the bulge and the disc
separately.
Therefore, we have 3 spectra to fit: one for the entire system, one
for the disc, and one for the bulge.

The fit is performed using the pPXF routine with a set of simple
stellar population (SSP) models from \citet{Vazdekis+10} and Gaussian
functions to remove emission lines as templates. The SSP templates
span a wide range of ages (from 0.5 to 14 Gyrs), and they include
sub-solar ([Z/H]$=-0.25]$), solar ([Z/H]$=0.06]$), and super-solar
([Z/H]$=+0.40]$) metallicities \footnote{This choice is justified by
  the range of values obtained in Section \ref{sec:ssp}}. We used SSP
models and not the ELODIE stars because, contrarily to Section
\ref{sec:kinematics}, we are interested in finding the stellar
population and not a precise fit to the kinematics.

The pPXF fitting routine computes the best-fitting stellar template as a
linear combination of all the individual SSP models in the library,
assigning a weight to each template.  We divided each SSP model by its
median flux so that the weight assigned by pPXF to each SSP indicates
the contribution to the total light. By removing this normalization,
the weights indicate the contribution of that SSP model to the total
mass.

Figure \ref{fig:sfh} shows the distributions of best-fitting weights
as a function of age and metallicity for the observed spectrum (top),
the disc component (middle) and the bulge component (bottom). The left column
shows the contribution to the total light while the right column
shows the contribution to the stellar mass.  We clearly identify 3
main components, an old population ($\geq7$ Gyr), an intermediate-age
population ($\approx$ 3 Gyr), and a young population ($\leq$1 Gyr). It
is beyond the scope of the paper to interpret the full star formation
history of NGC 3521 and study in detail how mass built up over
time. This, in theory, can be derived from the shape of the histograms
in the right panels of Fig. \ref{fig:sfh}. However, the fitting
procedure is sensitive to certain features in the spectra that do not
vary linearly with ages or with $\log$(age). For example, for age < 1
Gyr, the light (and therefore the features in the observed spectrum)
is dominated by blue massive stars, while for age > 5 Gyr, the light
is dominated by stars in the red giant branch. Therefore, one has to
be cautious in how to interpret the shape of the histograms in terms
of mass assembly over time. We therefore limit our analysis to 3 age
bins: $<1$ Gyr, 2--5 Gyrs, and $> 7$ Gyrs.

\subsubsection{Radial profiles}
\label{sec:sfh_radial}

\begin{figure}
\vbox{
 \psfig{file=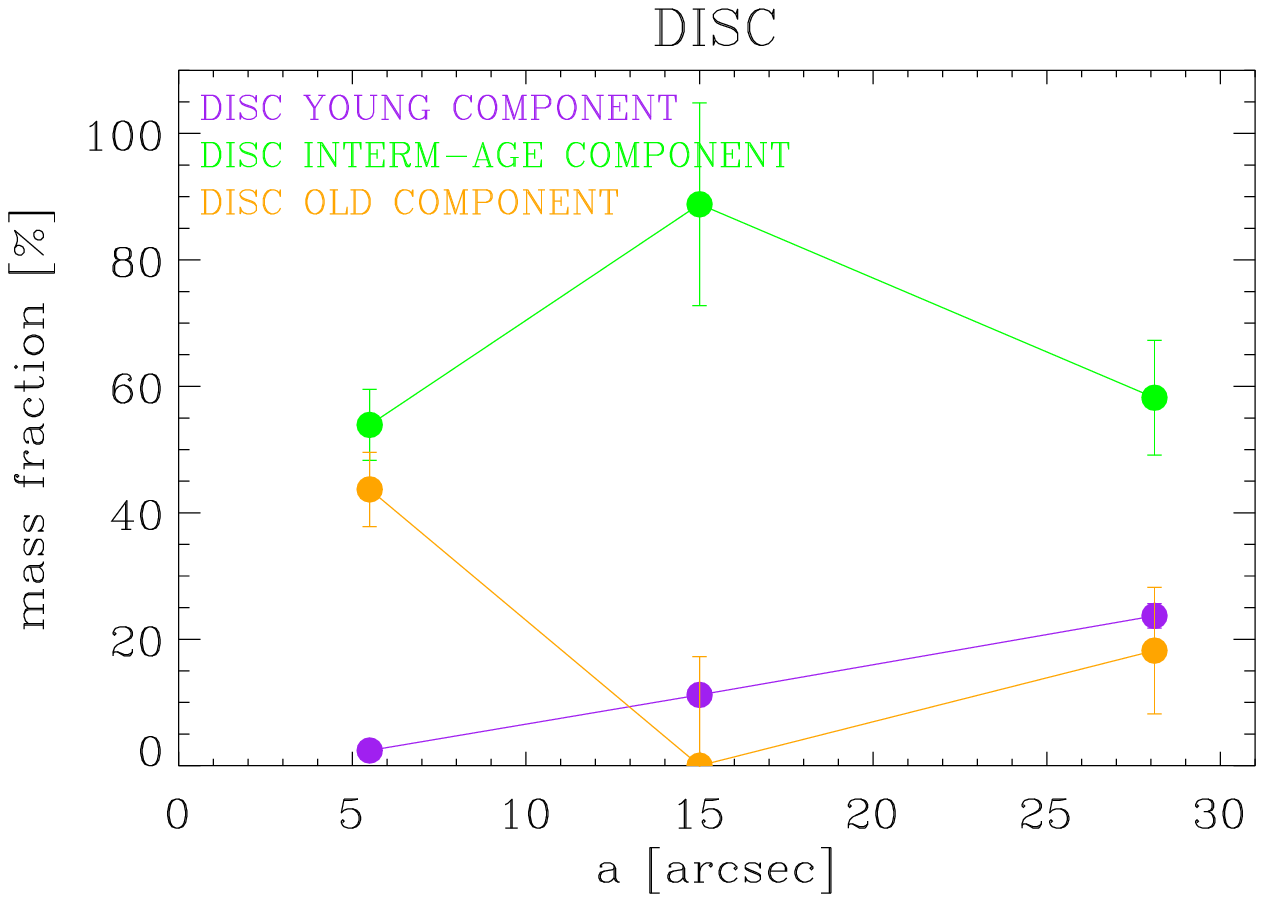,width=8.5cm}
 \psfig{file=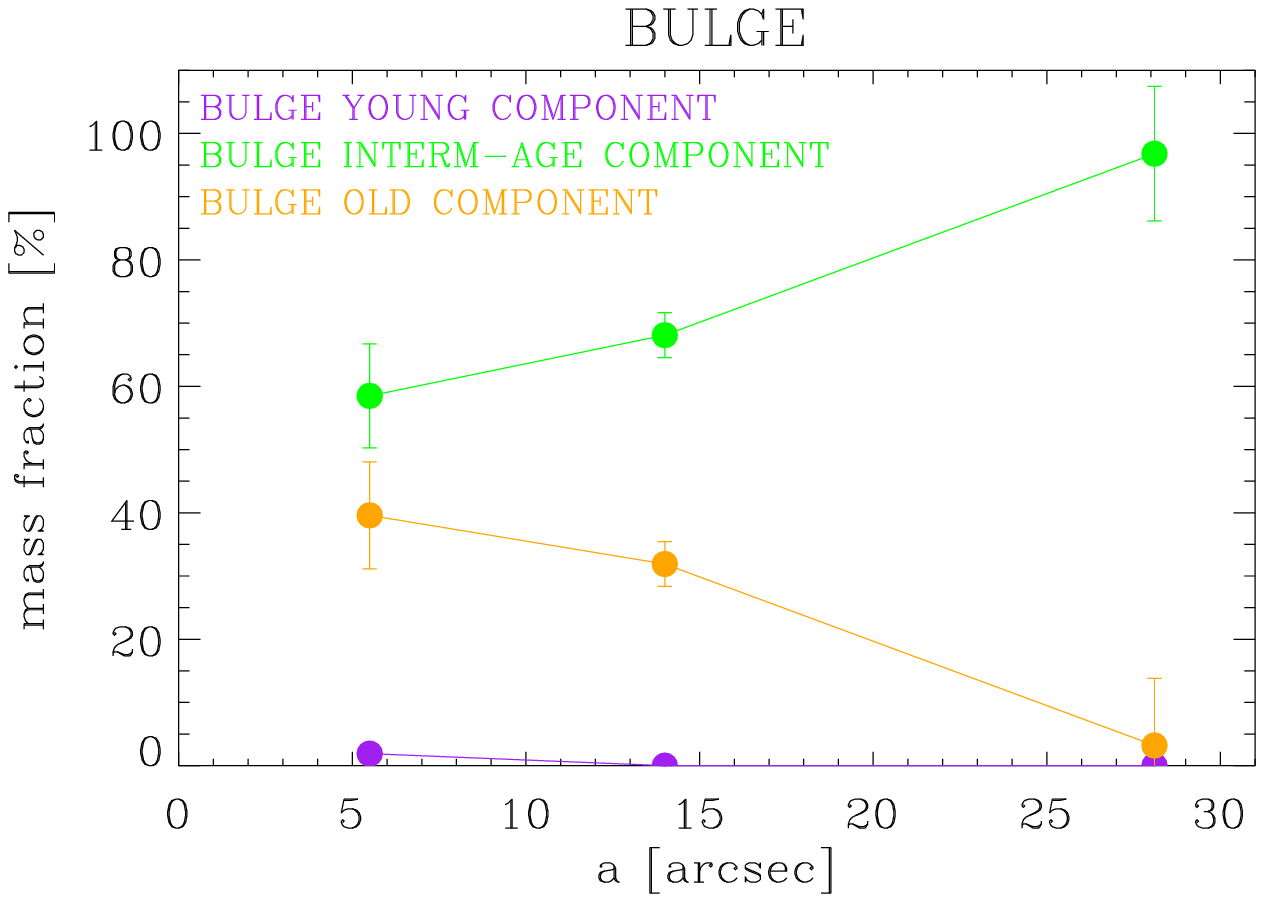,width=8.5cm}
}
\caption{Radial distribution of the contribution of each stellar
  population to the mass fraction of the disc (upper panel) and bulge (lower
  panel). }
\label{fig:sfh_radial}
\end{figure}

We repeated the analysis of multiple stellar populations by grouping
the spectra of the bulge and disc in three radial bins. A finer radial
description is beyond the signal-to-noise of our data.  The aim is to
study the radial variation of the contribution of the young,
intermediate, and old stellar populations to the mass of the bulge and
disc. This result is shown in Figure \ref{fig:sfh_radial}. The mass
fraction of the bulge and disc is dominated by the intermediate-age
component. The contribution of the young component to the disc mass
increases with radius, whereas it is negligible in the bulge.

\subsubsection{Computation of the errors in the contribution of each
  stellar population}
\label{sec:sfh_errors}

In order to compute the errors on the contribution to the light and
mass of each stellar population (Figs \ref{fig:sfh} and
\ref{fig:sfh_radial}, we used a boot-strapping Monte Carlo approach.

The fit of each spectrum was repeated 100 times using the best-fit
model with the addition of random noise as the input spectrum. The added
noise is consistent with the observed signal-to-noise of the fitted
spectrum. Each of these 100 realization gave a different result; the
standard deviation of the results gives us the estimate of the error
due to the signal-to-noise.

\subsection{Spatially-resolved simple stellar population}
\label{sec:ssp}

In this section we exploit the information of several absorption line
indices to infer the luminosity-weighted simple stellar population
properties (SSP) and their spatial distribution. This is done both for
the bulge and the disc. Because we fit simple stellar population
models to components that are know to host multiple stellar
populations (see Section \ref{sec:sfh}), the results are biased towards the
most ``important'' component. According to \citet{Serra+07}, the
luminosity-weighted stellar SSP age derived for multiple populations
will be biased towards the youngest and most luminous component. On
the other hand, the luminosity-weighted SSP metallicity is less
affected by this bias than the age, and therefore it traces better the
properties of the main bulk of stars.

For each spatial bin, we determined the best fitting linear
combination of ELODIE spectra (broadened to the spectral resolution of
VIRUS-W) for each of the kinematic components. After convolving these
best-fitting spectra to the LICK spectral resolution (FWHM=8.4 \AA),
we measured the line strengths of the indices \hbeta, \mgb, \feii, and
\feiii. The derived line strengths are then corrected using a linear
calibration to the Lick system that we obtained by comparing the
values measured on the stars in the stellar library that are in common
with the sample of \citet{Worthey+94}.  We then fit stellar population
models by \citet{Thomas+03} that account for variable
$\alpha$-elements abundance ratio to the measured indices.  We did not
include \fei\ in our procedure because it was the index that deviated
most from the best fit. The distribution of absolute normalized
residuals of \fei\ was about 4 times higher than those of the other
indices. Moreover, contrary to the other iron indices, the inclusion
of \fei\ in the fit systematically increased the global metallicity by
0.2 dex.

\begin{figure*}
\psfig{file=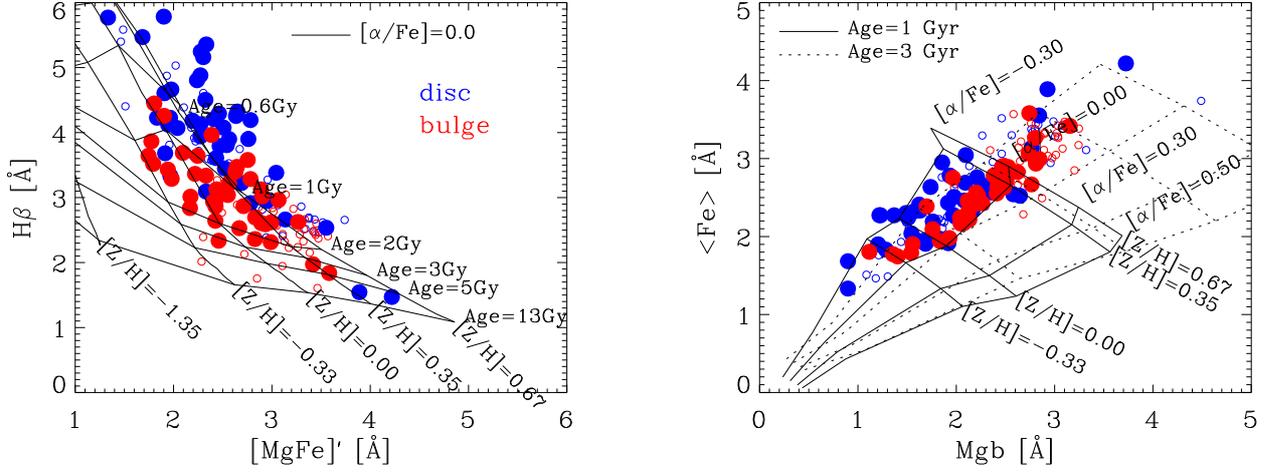,width=18cm,clip=}
\caption{Equivalent width of some of the absorption line indices
  measured in the bulge (blue) and disc (red) spectra produced by the
  spectral decomposition. Open symbols are the measurements in all the
  spatial bins, filled symbols are the measurements in those spatial
  bins were the velocity separation between disc and bulge is larger
  than 50 \kms. Predictions from single stellar population models by
  \citet{Thomas+03} are superimposed.}
\label{fig:diagnostic_plots}
\end{figure*}

For visualization purposes, in Figure \ref{fig:diagnostic_plots} we
show a combination of some of the measured indices in the \hbeta\
vs. \mgfe\ and \mgb\ vs \fem\ parameter spaces, and compare them with
the prediction of simple stellar population models.

\begin{figure*}
\psfig{file=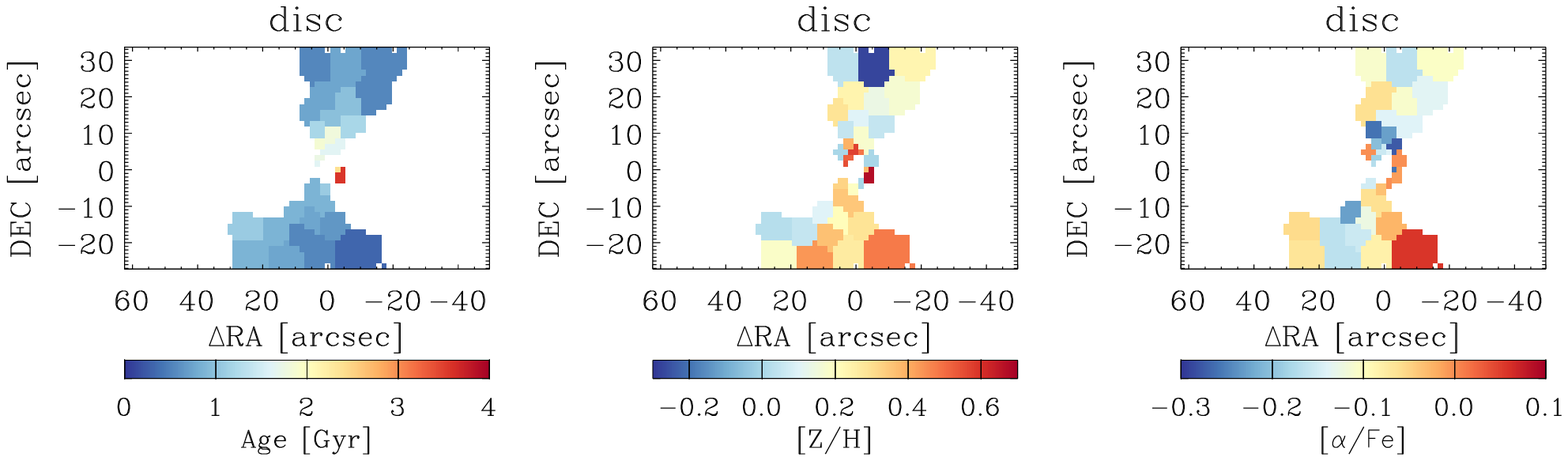,width=18cm,clip=,bb=10 385 588 515} 
\psfig{file=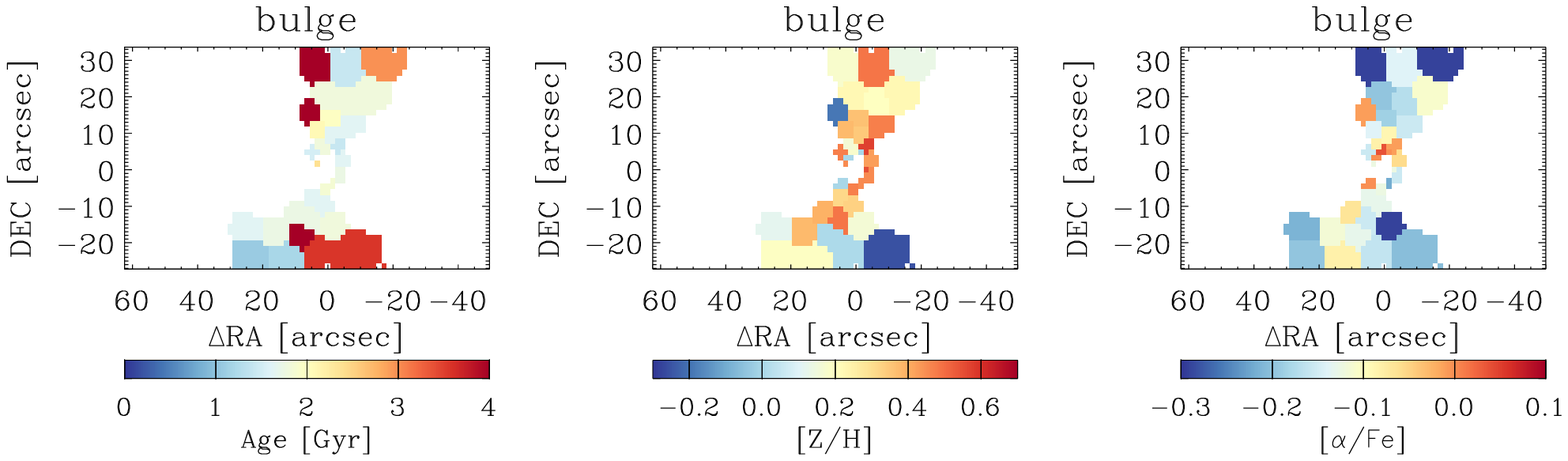,width=18cm,clip=,bb=10 385 588 515} 
\psfig{file=ssp_maps1.ps,width=18cm,clip=,bb=10 10 588 30}   
\psfig{file=ssp_maps1.ps,width=18cm,clip=,bb=10 342 588 385} 
\caption{Two-dimensional maps of age (left), metallicty (middle) and
  $\alpha$-enhancement for the disc (upper panels) and bulge (lower
  panels) of NGC\,3521. Only spatial bins outside the central 5'' and
  where the velocity separation between the two components is larger
  than 50 \kms\ are shown.}
\label{fig:ssp_maps}
\end{figure*}

\begin{figure}
 \psfig{file=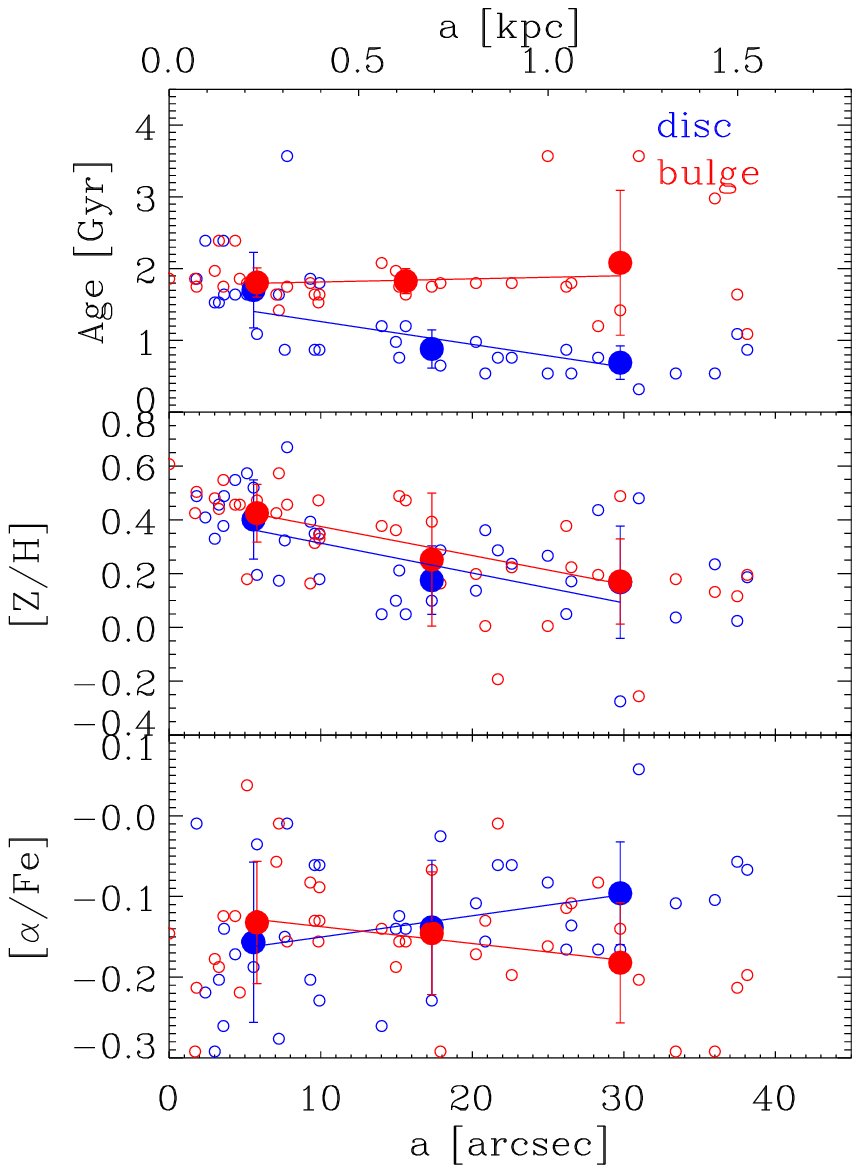,width=8cm,clip=}
 \caption{Values of age, metallicity and $\alpha$-enhancement for the
   disc (blue symbols) and bulge (red symbols) of NGC\,3521. Only bins
   where the velocity separation between the two components is larger
   than 50 \kms\ are considered. Values computed in each individual
   bin and plotted against the semi-major axis of the ellipse passing
   on that bin. Semi-major axis are computed considering ellipses with
   position angle $-19^{\circ}$ and ellipticity $e=0.35$
   \citep{Fabricius+12}. Small open symbols represent the values in
   individual bins, filled circles represent the mean value computed
   on elliptical annuli. Error bars of filled symbols are the standard
   deviations of the values within a give elliptical annulus. The
   continuous lines are the linear fit to the mean values.}
\label{fig:radial_profiles}
\end{figure}

The two-dimensional maps of ages, [Z/H], and [$\alpha$/Fe] are shown
in Figure \ref{fig:ssp_maps} and their radial profiles computed along
concentric ellipses oriented as the galaxy photometric major axis are
shown in Figure \ref{fig:radial_profiles}. Similarly to what we did in
Section \ref{sec:sfh}, we highlight in
Figs. \ref{fig:diagnostic_plots}-\ref{fig:radial_profiles} the spatial
bins where the absolute velocity difference between the two stellar
components is higher than 50 \kms.

\section{Discussion and conclusions}
\label{sec:discussion}

The spectroscopic decomposition clearly revealed the presence of two
stellar kinematic components in NGC\,3521, which we identified with
the bulge and the disc, and one ionized-gas component. The stars in
the bulge and disc co-rotate and are characterized by different
morphology, kinematics (velocity and velocity dispersion), stellar
population content, and equivalent width of the absorption lines.

The integrated fit of multiple stellar populations
(Sect. \ref{sec:sfh0}) reveals the presence of three populations of
stars that are highlighted in the upper panels of Fig.
  \ref{fig:sfh}:

\begin{itemize}
\item a young  ($\leq$1 Gyr) and metal poor population ([Z/H]$ \lesssim 0$);
\item an intermediate-age ($\approx$ 3 Gyr) and metal rich component
  ([Z/H]$ \gtrsim 0$), which dominates the light and mass of NGC
    3521;
\item an old population ($\geq7$ Gyr) that spans a large range of
  metallicity values ($-0.25 \lesssim $[Z/H]$ \lesssim 0.4$).
\end{itemize}

The presence of a young component (age $\leq 1$ Gyr, that we are able
to associate mainly with disc stars, see Fig. \ref{fig:sfh}) is
compatible with the independent detection of active star formation in
NGC\,3521 ($\log \Sigma_{SFH}/[{\rm M}_{\odot} yr^{-1} kpc^{-2}]$ =
-2.583, \citealt{Calzetti+10}).

Exploiting the spectral decomposition, we managed to remove the mutual
contamination of bulge and disc, and therefore we could investigate
how these stellar populations distribute among the bulge and the disc
of NGC\,3521. As shown in the middle and bottom panels of Figure
\ref{fig:sfh}, the light of the disc is dominated by the young and
intermediate populations, whereas the bulge is dominated by the
intermediate population. The mass of the disc is dominated by the
intermediate and old components, whereas the mass of the bulge is
dominated by the intermediate component. The stars of the old
population that are associated to the disc are metal rich
(Fig. \ref{fig:sfh}, middle panels), whereas the stars of the old
population that are associated to the bulge are metal poor
(Fig. \ref{fig:sfh}, bottom panels).

Figure \ref{fig:sfh_radial} indicates that the different populations
of stars have a distribution in the disc and bulge that depends on
radius. Indeed, the percentage of young stars in the disc increases
with radius. On the other hand, the young stars in the bulge are
concentrated only in the centre, and their (negligible) contribution
to the bulge mass decrease with radius.

The equivalent width of the spectral indices and the
luminosity-weighted simple stellar populations of the bulge are
different from those of the disc (Figures \ref{fig:diagnostic_plots},
\ref{fig:radial_profiles}). With respect to the luminosity-weighted
simple stellar population, the main difference between bulge and disc
is seen in their ages. The disc is younger and has a negative age
radial gradient (from $1.7 \pm 0.5$ Gyr in the innermost elliptical
bin down to $0.7 \pm 0.2$ Gyr in the outermost elliptical bin). The
bulge is slightly older with an average age of $1.9 \pm 0.4$ Gyr (see
Fig. \ref{fig:radial_profiles}).

Both components have very similar luminosity-weighed metallicity and
$\alpha$-enhancement, within the uncertainties. Their metal content is
super-solar and has a mild negative radial profile (from $0.4 \pm 0.1$
in the innermost elliptical bin down to $0.2 \pm 0.2$); their alpha
enhancement is sub-solar and nearly constant towards the entire
observed radial range ($\langle
$[$\alpha$/Fe]$\rangle \pm -0.14 \pm 0.03$).

The results of Sections \ref{sec:sfh0}, and \ref{sec:ssp} are
consistent with each other, at least from a qualitative point of
view. The range of luminosity-weighted ages as determined via the SSP
fit in Section \ref{sec:ssp} (disc $\leq 2$ Gyr, and bulge $\approx 2$
Gyr) is of the same order as the mean age of the luminosity-dominant
stellar populations measured in Section \ref{sec:sfh} ($\approx 3$
Gyr, Fig. \ref{fig:sfh} left panels). Moreover, the range of mean
luminosity-weighted metallicities ($0.2 \leq {\rm [Z/H]} \leq 0.4$,
Fig. \ref{fig:radial_profiles}) is consistent with the metallicity of
the main bulk of stars ($0 \leq {\rm [Z/H]} \leq 0.4$,
Fig. \ref{fig:sfh} right panels).  The negative age gradient of the
disc (Fig. \ref{fig:radial_profiles}, upper panel) is a consequence of
the mild radial increase of the contribution of young stars to the
mass of the disc (Fig. \ref{fig:sfh_radial}, upper panel).

From the combined information of the integrated multiple stellar
populations analysis (Sect. \ref{sec:sfh}), and the spatially-resolved
simple stellar population analysis (Sect. \ref{sec:ssp}), we can infer
the following formation scenario for NGC\,3521.

\begin{table}
\begin{center}
\begin{tabular}{l c c c}
\hline
Component &  $\Delta$Age      & $\Delta$[Z/H]        & $\Delta$[$\alpha$/Fe]  \\
          &  [Gyr kpc$^{-1}$]  & [dex kpc$^{-1}$]      & [dex kpc$^{-1}$]        \\
\hline
Disc      &  --0.8$\pm$ 0.4  & --0.3   $\pm$ 0.2    &  0.1  $\pm$ 0.1      \\   
Bulge     &    0.1 $\pm$ 0.1  & --0.27   $\pm$ 0.04     & --0.1 $\pm$ 0.1      \\
\hline
\end{tabular}
\end{center}
\caption{The measured radial gradient of the luminosity-weighted simple stellar population parameters for the disc and bulge components in NGC\,3521.}
\label{tab:gradients}
\end{table}

The galaxy formed with multiple formation episodes, the first
occurring $> 7$ Gyrs ago from metal rich material. A second episode of
star formation occurred about 3 Gyr ago; this episode involved both
components, but it affected mainly the bulge. Because the
intermediate-age population is also the most luminous, the age radial
profiles shown in Figure \ref{fig:radial_profiles} refer to this
second episode.

The third episode of star formation occurred $\leq 1.5$ Gyr ago and
 added mass mainly to the disc component. This population of stars dominates
the light of the disc component (although their mass contribution is
small), therefore the radial age gradient shown in Figure
\ref{fig:radial_profiles} describes these stars. This radial profile
suggests that the formation of the young disc stars started right after the
formation of the intermediate-bulge population of stars in the
innermost elliptical bin. Because the mass contribution of the
youngest disc star is very small ($\sim 8$\% of the disc stellar mass,
Fig. \ref{fig:sfh}), the radial metallicity gradient shown in Figure
\ref{fig:radial_profiles} describes the intermediate and old disc
stars.

The metallicity of the stars associated with the youngest disc component
ranges from solar to sub-solar (Fig. \ref{fig:sfh}), therefore
they cannot originate from recycled material of previous generation of
stars (either from disc or bulge), because the latter is are more metal
  rich. The material must come from outside the disc and the bulge of
NGC\,3521.  On the other hand, there is no evidence of gas accretion
onto NGC\,3521 from the intracluster medium \citep{Elson14}. The most
probable scenario is that the origin of this young component is
material coming from the so-called ``anomalous'' gas halo surrounding
NGC\,3521 detected by \citet{Elson14}. The authors also suggest that
the interplay between the stellar feedback and star formation is
regulated by galactic fountains that transfer material from and to
the halo, which is consistent with our interpretation.

The disc of NGC\,3521 has luminosity-weighted properties that are
similar to those of other spiral galaxies for which multiple
age-components are detected (e.g., \citealt{Morelli+15}). 
However, the majority of the disc light in the sample of
\citet{Morelli+15} is dominated by stars older than 7 Gyrs (Figure 2
in their paper), whereas the light of the disc of NGC 3521 is
dominated by stars younger than 7 Gyrs.

The bulge of NGC\,3521 is consistent with those of other bulges in
spirals where an attempt was made to minimize the contamination from
the disc \citep{Morelli+08, Morelli+12}. The majority of the bulges in
those studies have intermediate ages and super solar metallicity,
positive age gradients, and negative metallicity and $\alpha$
gradients, similar the bulge of NGC 3521.

The properties of NGC\,3521 differ from those of the majority of S0
galaxies in which spectral decomposition was attempted to isolate the
contribute of disc and bulge \citep{Johnston+12, Tabor+17}.  The
bulges in those S0 samples have intermediate ages, as the majority of
the stars in the bulge of NGC 3521. However there is evidence that the
bulges in those samples are younger than the corresponding discs,
unlike NGC 3521.

We cannot rule out merging as an alternative scenario opposite to the
in-situ formation of stars as builder of the structural components in
NGC\,3521, at least for the old and the intermediate
populations. Indeed, as discussed in Section \ref{sec:fors1}, the
galaxy shows deviations from a smooth axi-symmetric structure with
spiral arms. These structures could be the relics of the past merger
history of NGC\,3521 that contributed to the build-up of its
components. However, it is difficult to associate these non
axi-symmetric structures to the youngest disc component in NGC\,3521.
Indeed, they extend perpendicularly to the disc and are apparently
detached from the blue spiral structure that trace the youngest
component (Figure \ref{fig:fors1}). Therefore, if a merger indeed
contributed to the formation of the stellar components in NGC 3521, it
is more likely that it directly contributed to formation of the old
and/or the intermediate-age populations. The same merger could also be
the mechanism that allowed NGC 3521 to acquire the ``anomalous'' HI
cloud, from which the youngest stars in the disc have originated.

In summary, we propose the following formation scenario. The bulge and
disc components of NGC\,3521 formed a long time ago from stars of
similar age; the formation was followed by a rejuvenation process (or
accretion of younger stars via a merger) that involved both
components, but mainly the bulge. Then, very recently, the disc of NGC
3521 accreted material from the surrounding gas halo, which could have
been acquired during a past merger, and formed a new generation of
stars in an inside-out manner.

NGC 3521 is yet another good example of how it is possible to recover
the formation and mass assembly of bulges and discs via the
spectroscopic decomposition approach independently, especially for
cases of fainter bulges embedded in luminous discs, or vice-versa. The
next natural step for this kind of investigation is to study a
representative sample of galaxies to understand what are the preferred
formation channels.

\section*{Acknowledgements}

We wish to thank Michael Opitsch for useful discussion during the
preliminary phase of this work. LC thanks the Department of
Physics and Astronomy of the Padova University for hospitality while
this paper was in progress.

\bibliography{ngc3521}

\end{document}